\documentclass{vldb}
\usepackage{graphicx}
\usepackage{balance}  

\usepackage{microtype}

\usepackage{booktabs} 
\usepackage{verbatim}
\usepackage{makecell}
 \usepackage{mathtools}
\usepackage{color}
\usepackage{algorithm}
\usepackage{algorithmicx}
\usepackage{algpseudocode}
\usepackage{bbm}
\usepackage{tikz}
\usepackage{subfig}
\usepackage{hyperref}
\usetikzlibrary{arrows,automata,calc,shapes,patterns,decorations,positioning,shadows,intersections,3d,decorations.pathreplacing}
\usepackage{ntheorem}
\usepackage{thm-restate}
\usepackage[capitalise,noabbrev]{cleveref}
\usepackage[normalem]{ulem}
\usepackage{blkarray, bigstrut}
\usepackage{enumitem}

\newcommand{\revision}[1]{#1}

\newtheorem{thm}{Theorem}
\newtheorem{theorem}{Theorem}[section]
\newtheorem{proposition}[theorem]{Proposition}
\newtheorem{corollary}[theorem]{Corollary}

\newtheorem{definition}[theorem]{Definition}
\newtheorem{problem}[theorem]{Problem}
\newtheorem{example}[theorem]{Example}

\newtheorem{remark}[thm]{Remark}

\algnewcommand{\LeftComment}[1]{\Statex \(\triangleright\) #1}

\newcommand{\norm}[1]{\left\lVert#1\right\rVert}
\DeclareMathOperator*{\argmin}{arg\,min}

\newcommand*\X{\mathbf{X}}
\newcommand*\W{\mathbf{W}}
\newcommand*\V{\mathbf{V}}
\newcommand*\Q{\mathbf{Q}}
\newcommand*\R{\mathbf{R}}
\newcommand*\D{\mathbf{D}_{\Q}}
\newcommand*\x{\mathbf{x}}
\newcommand*\y{\mathbf{y}}
\newcommand*\z{\mathbf{z}}
\newcommand*\q{\mathbf{q}}
\newcommand*\w{\mathbf{w}}
\renewcommand{\v}{\mathbf{v}}
\newcommand*\reals{\mathbb{R}}

\newcommand{\M}{\ensuremath{\mathcal{M}}}
\newcommand{\U}{\ensuremath{\mathcal{U}}}
\renewcommand{\O}{\ensuremath{\mathcal{O}}}

\begin{document}


\title{A workload-adaptive mechanism for linear queries under local differential privacy}
\numberofauthors{1} 
\author{
\alignauthor
Ryan McKenna \hfill Raj Kumar Maity \hfill Arya Mazumdar \hfill Gerome Miklau \\
       \affaddr{College of Information and Computer Sciences \\ University of Massachusetts Amherst}       \email{\{rmckenna, rajkmaity, arya, miklau\}@cs.umass.edu}
       }
\date{30 July 1999}

\maketitle

\begin{abstract}
We propose a new mechanism to accurately answer a user-provided set of linear counting queries under local differential privacy (LDP).  Given a set of linear counting queries (the workload) our mechanism automatically adapts to provide accuracy on the workload queries.  We define a parametric class of mechanisms that produce unbiased estimates of the workload, and formulate a constrained optimization problem to select a mechanism from this class that minimizes expected total squared error. We solve this optimization problem numerically using projected gradient descent and provide an efficient implementation that scales to large workloads. We demonstrate the effectiveness of our optimization-based approach in a wide variety of settings, showing that it outperforms many competitors, even outperforming existing mechanisms on the workloads for which they were intended.
\end{abstract}

\section{Introduction}

In recent years, Differential Privacy \cite{dwork2006calibrating} has emerged as the dominant approach to privacy and its adoption in practical settings is growing.  
Differential privacy is achieved with carefully designed randomized algorithms, called mechanisms.  
The aim of these mechanisms is to extract utility from the data while adhering to the constraints imposed by differential privacy.  
Utility is measured on a task-by-task basis, and different tasks require different mechanisms.  Utility-optimal mechanisms, or mechanisms that maximize utility for a given task under a fixed privacy budget, are still not known in many cases.

There are two main models of differential privacy: the central model and the local model.  In the central model, users provide their data to a trusted data curator, who runs a privacy mechanism on the dataset in its entirety. In the local model, users execute a privacy mechanism before sending it to the data curator.  Local differential privacy (LDP) offers a stronger privacy guarantee than central differential privacy, as it does not rely on the assumption of a trusted data curator.  For that reason, it has been embraced by several organizations like Google \cite{erlingsson2014rappor}, Apple \cite{thakurta2017learning}, and Microsoft \cite{ding2017collecting} for the collection of personal data from customers. While the stronger privacy guarantee is a benefit of the local model, it necessarily leads to greater error than the central model~\cite{dwork2014algorithmic}, which makes error-optimal mechanisms an important goal. 

Our focus is answering a workload of linear counting queries under local differential privacy.  Answering a query workload is a general task that subsumes other common tasks, like estimating histograms, range queries, and marginals.  Furthermore, the expressivity of linear query workloads goes far beyond these special cases, as it can include an arbitrary set of predicate counting queries. By defining the workload, the analyst expresses the exact queries they care about most, and their relative importance.  
There are several LDP mechanisms for answering particular fixed workloads, like histograms \cite{acharya2018,ye2018optimal,bassily2017practical,wang2017locally}, range queries \cite{cormode2019answering,wang2019answering}, and marginals \cite{cormode2018marginal,wang2019answering}.  These mechanisms were carefully crafted to provide accuracy on the workloads for which they were designed, but their accuracy properties typically do not transfer to other workloads.  Some LDP mechanisms are designed to answer an arbitrary collection of linear queries~\cite{bassily2018linear,edmonds2019power}, but they do not outperform simple baselines in practice.  



In this paper, we propose a new mechanism that automatically adapts in order to prioritize accuracy on a target workload.  Adaptation to the workload is accomplished by solving a numerical optimization problem, in which we search over an expressive class of unbiased LDP mechanisms for one that minimizes variance on the workload queries.  

Workload-adaptation ~\cite{hardt2012simple,li2010optimizing,mckenna2018optimizing} is a much more developed topic in the central model of differential privacy and has led to mechanisms that offer best-in-class error rates in some settings~\cite{dpbench}. Our work is conceptually similar to the Matrix Mechanism \cite{li2010optimizing,mckenna2018optimizing}, which also minimizes variance over a class of unbiased mechanisms. However, because the class of mechanisms we consider is different, the optimization problem is fundamentally different and requires a novel analysis and algorithmic solution. 
\revision{We thoroughly discuss the similarities and differences between these two mechanisms in \cref{sec:related}. }



\paragraph*{Contributions}

The paper consists of four main technical contributions. 
\begin{itemize}
\itemsep0em
\item We propose a new class of mechanisms, called \emph{workload factorization mechanisms}, that generalizes many existing LDP mechanisms, and we formulate an optimization problem to select a mechanism from this class that is optimally tailored to a workload (\cref{sec:approach}).  
\item We give an efficient algorithm to approximately solve this optimization problem, by reformulating it into an algorithmically tractable form (\cref{sec:algorithm}).
\item We provide a theoretical analysis which illuminates error properties of the mechanism and justifies the design choices we made (\cref{sec:theory}). 
\item In a comprehensive set of experiments we test our mechanism on a range workloads, showing that it consistently delivers lower error than competitors, by as much as a factor of $14.6$ (\cref{sec:experiments}).
\end{itemize}

\section{Background and Problem Setup}
In this section we introduce notation for the input data and query workload as well as provide basic definitions of local differential privacy. A full review of notation is provided in \cref{table:notation} of the Appendix. 


\subsection{Input Data and Workload}

Given a domain $\U$ of $n$ distinct user types 
,the input data is a collection of $N$ users $\langle u_1, \dots, u_N \rangle$, where each $u_i \in \U$. We commonly use a vector representation of the input data, containing a count for each possible user type:
\begin{definition}[Data Vector] \label{def:datavec}
The data vector, denoted by $\x$, is an $n$-length vector of counts indexed by user types $u \in \U$ such that:
\vspace{-0.5em}
$$ x_u = \sum_{j=1}^{N} \mathbbm{1}\{u_j=u\} \quad \forall u \in \U. $$
\end{definition}
In the local model, we do not have direct access to $\x$, but it is still useful to define it for the purpose of analysis. Below is a simple data vector one might obtain from a data set of student grades.  

\begin{example}[Student Data]
Consider a data set of student grades, where $\U = \{ A, B, C, D, F \} $.  Suppose $10$ students got an $A$, $ 20$ students got a $B$, $5$ students got a $C$ and no students got a $D$ or $F$.  Then the data vector would be:
$$ \x = \begin{bmatrix} 10 & 20 & 5 & 0 & 0 \end{bmatrix}^T $$
\end{example}

Linear counting queries have a similar vector representation, as shown in \cref{def:query}.

\begin{definition}[Linear Query] \label{def:query}
A linear counting query is an $n$-length vector $\w$ indexed by user types $u \in \U$, such that the answer to the query is $\w^T \x = \sum_{u \in \U} w_u x_u $.
\end{definition}

A workload is a collection $p$ linear queries $\w_1, \dots, \w_p \in \reals^n$ organized into a $p \times n$ matrix $\W$. Our goal is to accurately estimate answers to each workload query under local differential privacy, i.e., we want to privately estimate $ \W \x $. 

The most commonly studied workload is the so-called Histogram workload, which is represented by a $n\times n$ identity matrix.  A more interesting workload is given below:

\begin{example}[Prefix workload] \label{ex:prefix}
The prefix workload contains queries that compute the (unnormalized) empirical cumulative distribution function of the data, or the number of students that have grades $\geq A, \geq B, \geq C, \geq D, \geq F$.  
$$
\W = 
\begin{bmatrix}
1 & 0 & 0 & 0 & 0 \\
1 & 1 & 0 & 0 & 0 \\
1 & 1 & 1 & 0 & 0 \\
1 & 1 & 1 & 1 & 0 \\
1 & 1 & 1 & 1 & 1 \\
\end{bmatrix}
$$
\end{example}

The workload $\W$ is an input to our algorithm, reflecting the queries of interest to the analyst and therefore determining the measure of utility that will be used to assess algorithm performance. 
\revision{
In this setup, we make no assumptions about the structure or contents of $\W$, and allow it to be completely arbitrary, even including the same query multiple times or multiple linearly dependent queries.}


\subsection{Local Differential Privacy}

Local differential privacy (LDP) \cite{dwork2014algorithmic} is a property of a randomized mechanism $\M$ that acts on user data.  Instead of reporting their true user type, users instead report a randomized response obtained by executing $\M$ on their true input.  These randomized responses allow an analyst to learn something about the population as a whole, while still providing the individual users a form of plausible deniability about their true input.  The formal requirement on $\M$ is stated in \cref{def:ldp}.

\begin{definition}[Local Differential Privacy] \label{def:ldp}
A randomized mechanism $\M : \U \rightarrow \O$ is said to be $\epsilon$-LDP if an only if, for all $u, u' \in \U$ and all $S \subseteq \O$:
$$ \Pr[\M(u) \in S] \leq \exp{(\epsilon)} \Pr[\M(u') \in S] $$
\end{definition}

The output range $\O$ can vary between mechanisms.  In some simple cases, it is the same as the input domain $\U$, but it does not have to be.  Choosing the output range is typically the first step in designing a mechanism.  When the range of the mechanism is finite, i.e., $|\O| = m$, we can completely specify a mechanism by a so-called \emph{strategy matrix} $\Q \in \reals^{m \times n}$, indexed by $(o,u) \in \O \times \U$. The mechanism $\M_{\Q}(u)$ is then defined by:
$$ \Pr[\M_{\Q}(u) = o] = Q_{o,u} $$

This encoding of a mechanism essentially stores a probability for every possible input, output pair in the strategy matrix $\Q$.  We translate Definition \ref{def:ldp} to strategy matrices in \cref{prop:strategy}.

\revision{
\begin{table*}[t]
\centering
\begin{tabular}{| c | c | c | l|} 
 \hline
 Mechanism &Input& Output & Strategy Matrix \\ 
 \hline 
 \hline
 \textbf{Randomized Response}~\cite{warner1965randomized}& $ u \in [n] $ &  $o \in [n] $ & $ Q_{o,u} \propto
\begin{cases} 
\exp{(\epsilon)} & o = u \\
1 & o \neq u 
\end{cases} $ \\
 \hline
 \textbf{RAPPOR}~\cite{erlingsson2014rappor}& \makecell {$u \in [n] $}& $ o \in \{0,1\}^n $ & $
Q_{o,u} \propto \exp{\Big(\frac{\epsilon}{2}\Big)}^{n - \norm{o - e_u}_1}
$\\
\hline
\textbf{Hadamard}~\cite{acharya2018} & $u \in [n]$ & \makecell {$o \in [K] $ \\$K=[2^{\lceil\log_2(n+1)\rceil}]$}&  $ Q_{o,u} \propto
\begin{cases} 
\exp{(\epsilon)} & H_{o+1,u} = 1  \\
1 & H_{o+1,u} = -1 \\
\end{cases} $ \\
\hline
\textbf{Subset Selection}~\cite{ye2018optimal} & $u \in [n]$ & \makecell {$ o\in \{0,1\}^n$ \\ $\norm{o}_{1}=d $} &$ Q_{o,u} \propto
\begin{cases} 
\exp{(\epsilon)} & o_u = 1 \\
1 &  o_u = 0
\end{cases} $ \\
\hline
\end{tabular}
\caption{Existing LDP mechanisms encoded as a strategy matrix. 
$e_u$ is the one-hot encoding of $u$, $\mathbf{H}$ is the $K \times  K$ Hadamard matrix, and $d$ is a hyper-parameter. \vspace{-1.5em}}
\label{table:stdmech}
\end{table*}
}

\begin{proposition}[Strategy Matrix] \label{prop:strategy}
The mechanism $\M_{\Q}$ is $\epsilon$-LDP if and only if the following conditions are satisfied:
\begin{enumerate}
\item $ Q_{o,u} \leq \exp{(\epsilon)} Q_{o,u'}, \:\: \forall \:\: o,u, u' .$
\item $ Q_{o,u} \geq 0, \forall o,u \:,\: \sum_{o} Q_{o,u} = 1, \forall \:\: u .$
\end{enumerate}
\end{proposition}
Above, the first condition is the privacy constraint, ensuring that the output distributions for any two users are close, and the second is the probability simplex constraint, ensuring that each column of $\Q$ corresponds to a valid probability distribution. 
Representing mechanisms as matrices is useful because it allows us to reason about them mathematically with linear algebra \cite{kairouz2014extremal,holohan2017extreme}.  
\cref{example:rr} shows how a simple mechanism, called randomized response\footnote{the name of this mechanism should not be confused with the outputs of an arbitrary mechanism $\M$, which we also call randomized responses.}, can be encoded as a strategy matrix.


\begin{example}[Randomized Response] \label{example:rr}
The randomized response mechanism \cite{warner1965randomized} can be encoded as a strategy matrix in the following way:
\vspace{-0.5em}
$$
\Q= \frac{1}{e^{\epsilon}+n-1}
\begin{bmatrix}
    e^{\epsilon} & 1 & \dots  & 1 \\
    1 & e^{\epsilon}  & \dots  & 1 \\
    \vdots & \vdots & \ddots & \vdots \\
    1 & 1 & \dots  & e^{\epsilon}
\end{bmatrix}
$$
\end{example}

For this mechanism, the output range is the same as the input domain, and hence the strategy matrix is square.  The diagonal entries of the strategy matrix are proportional to $e^{\epsilon}$, and the off-diagonal entries are proportional to $1$.  This means that each user reports their true input with probability proportional to $ e^{\epsilon}$ and all other possible outputs with probability proportional to $1$.  It is easy to see that the conditions of \cref{prop:strategy} are satisfied. While this is one of the simplest mechanisms, many other mechanisms can also be represented in this way. \revision{

For example, \cref{table:stdmech} shows how RAPPOR \cite{erlingsson2014rappor}, Subset Selection \cite{ye2018optimal} and Hadamard \cite{acharya2018hadamard} can be expressed as a strategy matrix.  Other mechanisms with more sophisticated structure, such as Hierarchical \cite{cormode2019answering,wang2019answering} and Fourier \cite{cormode2018marginal}, can also be expressed as a strategy matrix, but they require too much notation to explain in \cref{table:stdmech}.
A strategy matrix is simply a direct encoding of a conditional probability distribution, where every probability is explicitly enumerated as an entry of the matrix.  Hence, the representation can encode \emph{any} LDP mechanism, as long as the $\U$ and $\O$ are both finite and the conditional probabilities can be calculated.
}

When executing the mechanism, each user reports a (randomized) response $o_i = \M_{\Q}(u_i)$.  When all users randomize their data with the same mechanism, these responses are typically aggregated into a response vector $\y \in \mathbb{R}^m$  (indexed by elements $o \in \O$), where $y_o = \sum_{j=1}^{N} \mathbbm{1}\{o_j=o\} $.  \revision{Much like the data vector, the response vector is essentially a histogram of responses, as it counts the number of users who reported each response.  In the sequel, it is useful to think of the mechanism $\M_{\Q}$ as being a function from $\x$ to $\y$ instead of $u_i$ to $o_i$.  Thus,}
for notational convenience, we overload the definition of $\M_{\Q}$, allowing it to consume a data vector and return a response vector, so that $\M_{\Q}(\x) = \y$.  


The response vector $\y$ is often not that useful by itself, but it can be used to estimate more useful quantities, such as the data vector $\x$ or the workload answers $\W \x$.  This is typically achieved with a post-processing step, and does not impact the privacy guarantee of the mechanism. 



\vspace{-1em}
\section{The Factorization Mechanism} \label{sec:approach}

In this section, we describe our mechanism and the main optimization problem that underlies it.  
We begin with a high-level problem statement, and reason about it analytically until it is in a form we can deal with algorithmically.   We present our key ideas and the main steps of our derivation, but defer the finer details of the proofs to \cref{sec:pf}.  

Our goal is to find a mechanism that has low expected error on the workload.  This objective is formalized in \cref{prob:optimization}.
\vspace{-1em}
\begin{problem}[Workload Error Minimization] \label{prob:optimization}
Given a workload $\W$, design an $\epsilon$-LDP mechanism $\mathcal{M^*}$ that minimizes worst-case expected $L_2$ squared error.  Formally,
$$ \mathcal{M}^* = \argmin_{\mathcal{M}} \big\{ \max_{\x} \mathbb{E}[ \norm{\W \x - \mathcal{M}(\x)}_2^2] \big\}$$
\end{problem}
\vspace{-0.5em}
In the problem statement above, our goal is to search through the space of $\emph{all}$ $\epsilon$-LDP mechanisms for the one that is best for the given workload.  Because it is difficult to characterize an arbitrary mechanism $\M$ in a way that makes optimization possible, we do not solve the above problem in its full generality.  Instead, we perform the search over a restricted class of mechanisms which is easier to characterize.  While somewhat restricted, this class of mechanisms is quite expressive, and it captures many of the state-of-the-art LDP mechanisms available today \cite{erlingsson2014rappor,ye2018optimal,acharya2018hadamard,cormode2018marginal,cormode2019answering,wang2019answering}.

\begin{definition}[Workload Factorization Mechanism] \label{def:factorization}
Given an $\epsilon$-LDP strategy matrix $\Q \in \mathbb{R}^{m \times n}$ and a reconstruction matrix $\V \in \mathbb{R}^{p \times m}$ such that $\W = \V \Q$, the Workload Factorization Mechanism (factorization mechanism for short) is defined as:
\vspace{-0.5em}
$$ \mathcal{M}_{\V,\Q}(\x) = \V \mathcal{M}_{\Q}(\x) $$
\end{definition}
\vspace{-0.5em}
\revision{
Note that $\M_{\V,\Q}$ is defined in terms of $\M_{\Q}$, and it is parameterized by an additional \emph{reconstruction matrix} $\V$ as well.  This reconstruction matrix is used to estimate the workload query answers from the response vector output by $\M_{\Q}$.  
In fact, the workload query estimates produced by this class of mechanisms is unbiased, as:
$$\mathbb{E}[\M_{\V,\Q}(\x)] = \V \mathbb{E}[\M_{\Q}(\x)] = \V \Q \x = \W \x .$$
Furthermore, $\M_{\V,\Q}$ inherits the privacy guarantee of $\M_{\Q}$ by the post-processing principle of differential privacy~\cite{dwork2014algorithmic}.}


\begin{remark}[Consistency] \label{remark:extension}
Because these mechanisms are unbiased, they will produce the correct workload answers in expectation.  However, the individual estimates produced by the mechanism may not be the true workload answers for \emph{any} underlying dataset, which is a consistency problem.  For example, the estimates might suggest that one or more entries of the data vector are negative, which is clearly impossible.  To address this problem we show our mechanism can be improved through a post-processing technique that produces consistent estimates to the workload queries that are as close as possible to the unbiased estimates. This idea is not new, but an adaptation of existing techniques \cite{nikolov2013geometry,li2015matrix}, so we defer the full description to the appendix.  We nevertheless demonstrate its benefits experimentally in \cref{sec:exp:extension}.

We consider this an extension of our mechanism that can improve utility in practice, and evaluate it in isolation.  For the remainder of the paper, we will focus on the original, unbiased mechanism, as it is substantially easier to reason about analytically.
\end{remark}
Many existing LDP mechanisms can be represented as a factorization mechanism.  For example, we show how the Randomized Response mechanism can be expressed as a factorization mechanism in \cref{example:rr2}.

\begin{example}[Randomized Response] \label{example:rr2}
The randomized response mechanism uses $\Q$ as defined in \cref{example:rr} and $\V = \Q^{-1}$ to estimate the Histogram workload ($\W = \mathbf{I}$).
$$
\V= \frac{1}{e^{\epsilon}-1}
\begin{bmatrix}
    e^{\epsilon}+n-2 & -1 & \dots  & -1 \\
    -1 & e^{\epsilon}+n-2 & \dots  & -1 \\
    \vdots & \vdots & \ddots & \vdots \\
    -1 & -1 &  \dots  & e^{\epsilon}+n-2
\end{bmatrix}
$$
\end{example}

While the randomized response mechanism is intended to be used to answer the Histogram workload, there is no reason why it cannot be used for other workloads as well.  In fact, it is quite straightforward to see how it can be extended to answer an arbitrary workload, simply by using $\V = \W \Q^{-1} $.  

\subsection{Variance Derivation}

While the factorization mechanism is unbiased for any workload factorization, different factorizations lead to different amounts of variance on the workload answers.  This creates the opportunity to choose the workload factorization that leads to the lowest possible total variance. In order to do that, we need an analytic expression for the total variance in terms of $\V$ and $\Q$, which we derive in \cref{thm:variance}.  

\begin{restatable}[Variance]{theorem}{variance} \label{thm:variance}
The expected total squared error (total variance) of a workload factorization mechanism is:
\begin{align*}
\mathbb{E}[ \norm{\W \x - \mathcal{M}_{\V,\Q}(\x)}_2^2] &= 
\sum_{u \in \U} x_u \sum_{i=1}^p \v_i^T \text{Diag}(\q_u) \v_i - (\v_i^T \q_u)^2
\end{align*}
where $\q_u$ denotes column $u$ of $\Q$ and $\v_i^T$ denotes row $i$ of $\V$.
\end{restatable} 

Notice above that the exact expression for variance depends on the data vector $\x$, which we do not have access to, as it is a private quantity.  We want our mechanism to work well for all possible $\x$, so we consider \emph{worst-case} variance and a relaxation \emph{average-case} variance instead.\footnote{Alternatively, if we had a prior distribution over $\x$, we could use that to estimate variance.}
\begin{corollary}[Worst-case variance] \label{thm:worst-variance}
The worst-case variance of $\M_{\V,\Q}$ occurs when all users have the same worst case type (i.e., $x_u = N$ for some $u$), and is:
$$L_{worst}(\V,\Q) = N \max_{u \in \U} \sum_{i=1}^p \v_i^T \text{Diag}(\q_u) \v_i - (\v_i^T \q_u)^2. $$
\end{corollary}
\begin{corollary}[Average-case variance]
The average-case variance of $\M_{\V,\Q}$ occurs when $x_u = \frac{N}{n}$ for all $u$ and is:
$$L_{avg}(\V,\Q) = \frac{N}{n} \sum_{u \in \U} \sum_{i=1}^p \v_i^T \text{Diag}(\q_u) \v_i - (\v_i^T \q_u)^2.$$
\end{corollary}

With these analytic expressions for variance, we can analyze and compare existing mechanisms that can be expressed as a workload factorization mechanism.  The variance for randomized response is shown in \cref{example:varrr}.

\begin{example}[Variance of Randomized Response] \label{example:varrr}
The worst-case and average-case variance of the factorization in \cref{example:rr2}
on the Histogram workload is:
$$ L_{worst}(\V,\Q) = L_{avg}(\V,\Q) = N (n-1) \Big[ \frac{n}{(e^\epsilon-1)^2} + \frac{2}{e^\epsilon-1} \Big] $$
\end{example}

The expression above is obtained by simply plugging in $\V$ and $\Q$ to the equations above and simplifying.  Interestingly, the worst-case and average-case variance are the same for this workload factorization due to the symmetry in the workload and strategy matrices.

\subsection{Strategy Optimization} \label{sec:strategyoptimization}

With an analytic expression for variance, we can state the optimization problem underlying the factorization mechanism.  Our goal is to find a workload factorization that minimizes the total variance on the workload.  To do that, we set up an optimization problem, using total variance as the objective function while taking into consideration the constraints that have to hold on $\V$ and $\Q$.  This is formalized in \cref{prob:factorization}. 

\begin{problem}[Optimal Factorization] \label{prob:factorization}
Given a workload $\W$ and a privacy budget $\epsilon$:
\begin{equation*}
\begin{aligned}
& \underset{\V,\Q}{\text{minimize}}
& & L(\V,\Q) \\
& \text{subject to}
& & \W = \V \Q \\
& & & \textstyle \sum_{o} Q_{o,u} = 1 \quad \forall u \\
& & & 0 \leq Q_{o,u} \leq \exp{(\epsilon)} Q_{o,u'} \quad \forall o, u, u' .
\end{aligned}
\end{equation*}
\end{problem}

Above, $L$ is a loss function that captures how good a given factorization is, such as the worst-case variance $L_{worst}$ or the average-case variance $L_{avg}$.  
While our original objective in \cref{prob:optimization} was to find the mechanism that optimizes worst-case variance, for practical reasons we use the average-case variance instead. The average-case variance is a smooth approximation of the worst-case variance, which leads to a more analytically and computationally tractable problem.  Additionally, the smoothness of the average-case variance makes the corresponding optimization problem more amenable to numerical methods.  We study the ramifications of this relaxation theoretically in \cref{sec:worstvsavg1}.
%
%
When using $L_{avg}$ as the objective function, we observe it can be expressed in a much simpler form using matrix operations, as shown in \cref{thm:objective}.  

\begin{restatable}[Objective Function]{theorem}{objective} \label{thm:objective}
The objective function $L(\V,\Q) = tr[\V \D \V^T]$ is related to $L_{avg}(\V,\Q)$ by:
$$ L_{avg}(\V,\Q) = \frac{N}{n} \big(L(\V,\Q) - \norm{\W}_F^2\big), $$

where $\D = \text{Diag}(\Q \mathbf{1})$ and $tr[\cdot]$ is the trace of a matrix.
\end{restatable}

From now on, when we refer to $L(\V,\Q)$, we are referring to its definition in \cref{thm:objective}.  The new objective is equivalent to $L_{avg}$ up to constant factors, and hence can be used in place of it for the purposes of optimization. 

With this simplified objective function, we observe that for a fixed strategy matrix $\Q$, we can compute the optimal $\V$ in closed form.  If the entries of the response vector were statistically independent and had equal variance, then this would simply be $\V = \W \Q^\dagger$, where $\Q^\dagger$ is the Moore-Penrose pseudo-inverse of $\Q$ \cite{casella2002statistical,hay2010boosting}.  
However, since the entries of the response vector have unequal variance and are not statistically independent in general, this simple expression is not correct. 
We can still express the optimal $\V$ in closed form, however, as shown in \cref{thm:optv}.

\begin{restatable}[Optimal $\V$ for fixed $\Q$]{theorem}{optv} \label{thm:optv}
For a fixed $\Q$, the minimizer of $L(\V, \Q)$ subject to $\W = \V \Q$ is given by:
$$ \V = \W (\Q^T \D^{-1} \Q)^{\dag} \Q^T \D^{-1}, $$
\end{restatable}

Note that we can assume $\D$ is invertible without loss of generality.  If it were not, then one entry of the diagonal would have to be $0$, implying that a row of $\Q$ is all zero.  Such a row corresponds to an output that never occurs under the privacy mechanism, and can be removed without changing the mechanism.  Further note that for the above formula to apply, there must exist a $\V$ such that $\W = \V \Q$, which is guaranteed if and only if $\W$ is in the row space of $\Q$ \cite{strang1993introduction}.  Expressed as a constraint, this is $\W = \W \Q^\dag \Q $. 


Now that we know the optimal $\V$ for any $\Q$, we can plug it into $L(\V,\Q)$ to express the objective as a function of $\Q$ alone.  Doing this, and simplifying further, leads to our final optimization objective, stated in \cref{thm:objfunQ}.

\begin{restatable}[Objective Function for $\Q$]{theorem}{objfunQ} \label{thm:objfunQ}
The objective function can be expressed as:
$$L(\Q) = tr[(\Q^T \D^{-1} \Q)^{\dag} (\W^T \W)]. $$
\end{restatable}

$L(\Q)$ is our final optimization objective.  We are almost ready to restate the optimization problem in terms of $\Q$.  However, before we do that, it is useful to simplify the constraints of the problem.  
The constraints stated in \cref{prob:factorization} are challenging to deal with algorithmically because there are a large number of them.  Ignoring the factorization constraint, there are $n^2 m + n$ constraints on $\Q$, and each entry of $\Q$ is constrained by entries from the same column \emph{and} entries from the same row.  

By introducing an auxiliary optimization variable $\z \in \reals^m$, we reduce this to $n m + n$ constraints, so that each entry of $\Q$ is only constrained by entries from the same column and $\z$.  
Specifically, $\z$ corresponds to the minimum allowable value on each row of $\Q$, and every column of $\Q$ must be between $\z$ and $\exp{(\epsilon)} \z$ (coordinate-wise inequality).  It is clear that this is exactly equivalent to Condition $2$ in \cref{prop:strategy}. Also note that Condition $1$ can be expressed in matrix form as $ \Q^T \mathbf{1} = \mathbf{1}$.  The final optimization problem underlying the workload factorization mechanism is stated in \cref{prob:strategyopt}.

\begin{problem}[Strategy Optimization] \label{prob:strategyopt}
Given a workload $\W$ and a privacy budget $\epsilon$: 
\begin{align*}
& \underset{\Q, \z}{\text{minimize}}
& & tr[(\Q^T \D^{-1} \Q)^{\dag} (\W^T \W) ] \\
& \text{subject to}
& & \W = \W \Q^\dag \Q \\
& & & \Q^T \mathbf{1} = \mathbf{1} \\
& & & \mathbf{0} \leq \z \leq \q_u \leq \exp{(\epsilon)} \z \quad \forall u.\\
\end{align*}
\end{problem}
\vspace{-2em}
%

\section{Optimization Algorithm} \label{sec:algorithm}

We now discuss our approach to solving \cref{prob:strategyopt}.  It is a nonlinear optimization problem with linear and nonlinear constraints.  While the objective is smooth (and hence differentiable) within the boundary of the constraint $\W = \W \Q^\dag \Q$, it is not convex.  
It is typically infeasible to find a closed form solution to such a problem, and conventional numerical optimization methods are not guaranteed to converge to a global minimum for non-convex objectives.  However, such numerical gradient-based methods have seen remarkable empirical success in a variety of domains, often finding high quality local minima. That is the approach we take, however, rather than use an out-of-the-box commercial solver, which would not be able to scale to larger problem sizes, we provide our own optimization algorithm which achieves greater scalability by exploiting structure in the constraint set.


The algorithm we propose is an instance of \emph{projected gradient descent} \cite{nocedal2006numerical}, a variant of gradient descent that handles constraints.  To implement this algorithm, the key challenge is to \emph{project} onto the constraint set.   In other words, given a matrix $\R$ that does not satisfy the constraints, find the ``closest'' matrix $\Q$ that does satisfy the constraints.  Ignoring the constraint $\W = \W \Q^\dag \Q$ for now, this sub-problem is stated formally in \cref{prob:projection}.  

\begin{algorithm}[t]                   
\caption{Projection onto bounded probability simplex}\label{alg:projection}         
\begin{algorithmic}                 
\State \textbf{Input:} $\mathbf{r}, \z \in \mathbb{R}^m, \epsilon $
\LeftComment{sorted vector and corresponding permutation}
\State $ \mathbf{u}, \boldsymbol{\pi} = \text{sort}([\z - \mathbf{r}, \exp{(\epsilon)} \z - \mathbf{r} ]) $ 
\State $ \mathbf{a} = \begin{bmatrix} (-1)^{\mathbbm{1}[\pi_i \leq m]} \end{bmatrix}_{i=1 \dots 2m} $
\State $ \mathbf{b} = \begin{bmatrix} \sum_{j=1}^{i-1} a_j \end{bmatrix}_{i=1 \dots 2m} $
\State $ \rho = \min \{ 1 \leq i \leq 2 m : \mathbf{1}^T \z + b_i u_i + \sum_{j=1}^{i-1} a_j u_j > 1 \} $
\State $ \lambda = (1 - \mathbf{1}^T \z - b_{\rho} u_{\rho} - \sum_{j=1}^{\rho-1} a_j u_j) / b_{\rho} + u_{\rho} $
\State \textbf{return} $\text{clip}(\mathbf{r} + \lambda, \z, \exp{(\epsilon)} \z)$
\end{algorithmic}
\end{algorithm}

\begin{problem}[Projection onto LDP Constraints] \label{prob:projection}
Given an arbitrary matrix $\R$, a vector $\z$, and a privacy budget $\epsilon$, the projection onto the privacy constraint, denoted $\Q = \Pi_{\z,\epsilon}(\R)$ is obtained by solving the following problem:
\begin{align*}
& \underset{\Q}{\text{minimize}}
& & \norm{\Q - \R}_F^2 \\
& \text{subject to}
& & \Q^T \mathbf{1} = \mathbf{1} \\
& & & \z \leq \q_u \leq \exp{(\epsilon)} \z \quad \forall u.
\end{align*}
\end{problem}
\cref{prob:projection} is easier to solve than \cref{prob:strategyopt} because the objective is now a quadratic function of $\Q$.
%
In addition, a key insight to solve this problem efficiently is to notice that it is closely related to the problem of projecting onto the probability simplex \cite{wang2013projection} (now with bound constraints), and admits a similar solution. Specifically, the form of the solution is stated in \cref{prop:projection}.  

\begin{proposition}[Projection Algorithm] \label{prop:projection}
The solution to \cref{prob:projection} may be obtained one column at a time using
$$ \q_u = \text{clip}(\mathbf{r}_u + \lambda_u, \z, \exp{(\epsilon)} \z), $$
where $\text{clip}$ ``clips'' the entries of $\mathbf{r}_u + \lambda_u $ to the range $[\z, \exp{(\epsilon)} \z]$ entry-wise
and $\lambda_u$ is a scalar value that makes $ \mathbf{1}^T \q_u = 1 $.
\end{proposition} 

The solution is remarkably simple.  Intuitively, we add the same scalar value to every entry of $\mathbf{r}_u$ then clip those values that lie outside the allowed range.  The scalar value $\lambda_u$ is \revision{the Lagrange multiplier on the constraint $ \mathbf{1}^T \q_u = 1$, and is chosen so that $\q_u$ satisfies the constraint.} 
It may be calculated through binary search or any other method to find the root of the function $ \mathbf{1}^T \q_u - 1 = 0$.  
We give an efficient $O(m \log{(m)})$ \revision{algorithm to find $\lambda_u$ and $\q_u$} in \cref{alg:projection}.

Now that we have discussed the projection problem and its solution, we are ready to state the full projected gradient descent algorithm for finding an optimized strategy.  \cref{alg:matmech} is an iterative algorithm, where in each iteration we perform a gradient descent plus projection step on the optimization parameters $\z$ and $\Q$.  The gradient $\nabla_{\Q} L$ is easily obtained as $L$ is a function of $\Q$, but the gradient term $\nabla_{\z} L$ is less obvious.  However, by observing that $\Q$ is actually a function of $\z$ (from the projection step $\Pi_{z,\epsilon}$), we can use the multi-variate chain rule to back-propagate the gradient from $\Q$ to $\z$ to obtain $\nabla_{\z} L$.  We do not discuss the details of computing the gradients here, as it can be easily accomplished with automatic differentiation tools \cite{griewank1989automatic,maclaurin2015autograd}.  

\begin{algorithm}[t]                   
\caption{Strategy optimization}\label{alg:matmech}         
\begin{algorithmic}                 
\State \textbf{Input:} Workload $\W \in \mathbb{R}^{p \times n}$, privacy budget $\epsilon$
      \State \textbf{Intialize} $\Q \in \mathbb{R}^{m \times n}, \z \in \mathbb{R}^m $, $\beta \in \mathbb{R}_+$
      \State $\alpha = \beta / n \exp{(\epsilon)} $
         \For{$t = 1, \dots, T$}
            \State $\z \leftarrow \text{clip}(\z - \alpha \nabla_{\z} L(\Q), \mathbf{0}, \mathbf{1})$
       \State  $ \Q \leftarrow \Pi_{\z, \epsilon} (\Q - \beta \nabla_{\Q} L(\Q)) $
          \EndFor
       \State \textbf{return}  $\Q$
\end{algorithmic}
\end{algorithm}

We note that \cref{alg:matmech} handles the constraint $\W = \W \Q^\dag \Q$ ``for free'' in the sense that we do not need to deal with it explicitly, as long as the step sizes and initialization are chosen appropriately.   Specifically, as long as the initial $\Q$ satisfies the constraint, and the step sizes are sufficiently small, every subsequent $\Q$ in the algorithm will also satisfy the constraint.  Intuitively, this is because as we move closer to the boundary of the constraint, the objective function blows up and eventually reaches a point of discontinuity when the constraint is not satisfied.  Because we update using the negative gradient, which is a descent direction, we will never approach the boundary of the constraint.  
We discuss the choice of step size and initialization below. 
This trick is a very nice way to deal with a constraint that is otherwise challenging to deal with.  We note that similar ideas have been used to deal with related constraints in prior work \cite{yuan2016convex}.

The step size for the gradient descent step must be supplied as input, and two different step sizes are used to update $\Q$ and $\z$.  Notice that we take a smaller step size to update $\z$ than $\Q$.  This is a heuristic we use to make sure $\z$ doesn't change too fast; it improves the robustness of the algorithm.  We perform a hyper-parameter search to find a step size that works well, only running the algorithm for a few iterations in this phase, then running it longer once a step size is chosen. Decaying the step size at each iteration is also possible, as smaller step sizes typically work better in later iterations. 

The final missing piece in \cref{alg:matmech} is the initialization of $\Q$, for which there are multiple options.  One option is to initialize with the strategy matrix from an existing mechanism, such as the best one from \cref{table:stdmech}.  Then intuitively the optimized strategy will never be worse than the other mechanisms, because the negative gradient is a descent direction.  This is an informal argument, as it is technically possible that the optimized strategy has better average-case variance but worse worst-case variance than the initial strategy.  We do not take this approach, however, as we find initializing $\Q$ randomly tends to work better.  Specifically, we let $\R$ be a random $ 4n \times n$ matrix, where each entry is sampled from $U[0,1]$.  Then we obtain $\Q$ by projecting onto the constraint set; i.e., $\Q = \Pi_{\z,\epsilon}(\R)$, where $\z = \frac{1+e^{-\epsilon}}{8n} \mathbf{1}$, where $\mathbf{1}$ is a vector of ones.  
The choice of $m$ is also an important consideration when initializing $\Q$.  While larger $m$ leads to a more expressive strategy space, it also leads to more expensive optimization.  
Our choice of $m = 4n$ represents a sweet spot that we found works well empirically \revision{across a variety of workloads.  In general, a hyper-parameter search can be executed to find the best $m$.  This hyper-parameter search does not degrade the privacy guarantee in any way because we can evaluate the quality of a strategy without consuming the privacy budget, by using the analytic formulas for variance.}

\revision{
It requires $O(n^2 m + n^3)$ time to evaluate the objective function and its gradient (assuming $\W^T \W$ has been precomputed) and $O(n m \log{m})$ time to perform the projection.
Thus, the per-iteration time complexity of \cref{alg:matmech} is $O(n^2 m + n^3 + n m \log{m})$, or $O(n^3)$ when using $m = 4 n$.  
} 


\section{Theoretical Results} \label{sec:theory}

In this section, we answer several theoretical questions about our mechanism. First, we justify the relaxation in the objective function, used to make the optimization analytically tractable.  Second, we theoretically analyze the error achieved by our mechanism, measured in terms of sample complexity.  Third, we derive lower bounds on the achievable error of workload factorization mechanisms. 
All the proofs are deferred to \cref{sec:pf}.
\subsection{Relaxed Objective} \label{sec:worstvsavg1}

In \cref{sec:approach} we replaced our true optimization objective $L_{worst}$ with a relaxation $L_{avg}$.  In this section, we justify that choice theoretically, showing that $L_{worst}$ is tightly bounded above and below by $L_{avg}$.  
\begin{restatable}[Bounds on $L_{worst}$]{theorem}{avgworst} \label{thm:avg_worst}
Let $\W = \V \Q$ be an arbitrary factorization of $\W$ where $\Q$ is an $\epsilon$-LDP strategy matrix. Then the worst case variance $L_{worst}(\V,\Q)$ and average-case variance $L_{avg}(\V,\Q)$ are related as follows:
\begin{align*}
L_{avg}(\V,\Q) \leq L_{worst}(\V,\Q) \leq e^{\epsilon}\big(L_{avg}(\V,\Q) +\frac{N}{n}||\W||_F^2\big)
\end{align*}
\end{restatable}
\cref{thm:avg_worst} suggests that relaxing $L_{worst}$ to $L_{avg}$ does not significantly impact the optimization problem.  
Intuitively, this theorem holds because of the privacy constraint on $\Q$, which guarantees that the column of $\Q$ for the worst-case user cannot be too different from any other column. Hence, all users must have a similar impact on the total variance of the mechanism. 
Empirically, we find that $L_{worst}$ is often even closer to $L_{avg}$ than the upper bound suggests.  
Furthermore, in some cases $L_{worst}$ is exactly equal to $L_{avg}$, as we showed in \cref{example:varrr}.

\subsection{Sample complexity}

We gave an analytic expression for the expected total squared error (total variance) of our mechanism in \cref{thm:worst-variance}.  However, this quantity might be difficult to interpret, and it is more natural to look at the number of samples needed to achieve a fixed error instead.  Furthermore, when running an LDP mechanism it is important to know how much data is required to obtain a target error rate, as that information is critical for determining an appropriate privacy budget.   

Because the total variance increases with the number of individuals $N$ and the number of workload queries $m$, we instead look at a normalized measure of variance.

\begin{definition}[Normalized Variance]
The normalized worst-case variance of $\M_{\V,\Q}$ is:
$$ L_{norm}(\V,\Q) =\max_{\x} \mathbb{E}\Big[\frac{1}{m} \norm{\frac{1}{N} \big\{ \W \x - \M_{\V,\Q}(\x) \big\}}_2^2 \Big] $$
\end{definition}

$L_{norm}$ is the same as $L_{worst}$ up to constant factors, although it is more interpretable because it is a measure of variance on a single ``average'' workload query, where variance is measured on the normalized data vector $\x/N$.  

\begin{corollary}[Normalized Variance] \label{cor:norm-variance}
The normalized variance is:
\begin{align*} 
L_{norm}(\V,\Q) &= \frac{1}{m N^2} L_{worst}(\V,\Q) \\
&= \frac{1}{m N} \max_{u \in \U} \sum_{i=1}^p \v_i^T \text{Diag}(\q_u) \v_i - (\v_i^T \q_u)^2
\end{align*}
\end{corollary}

Interestingly, the dependence on $N$ does not change with $\V$ and $\Q$ --- it is always $\Theta(\frac{1}{N})$, but the constant factor depends on the quality of the workload factorization.  

\begin{corollary}[Sample Complexity] \label{cor:sample-complexity} 
The number of samples needed to achieve normalized variance $\alpha$ is:
\vspace{-0.5em}
$$ N \geq \frac{1}{m \alpha} \max_{u \in \U} \sum_{i=1}^p \v_i^T \text{Diag}(\q_u) \v_i - (\v_i^T \q_u)^2 $$
\end{corollary}
We can readily compute the sample complexity numerically for any factorization $\V \Q$.  In fact, the sample complexity and worst-case variance of a mechanism are proportional, as evident from the above equation.  Additionally, by replacing $L_{worst}(\V,\Q)$ with a lower bound, we can get an analytical expression for the sample complexity in terms of the privacy budget $\epsilon$ and the properties of the workload $\W$.

\begin{example}[{\fontsize{7.5pt}{7.5pt}\selectfont Sample complexity, Randomized Response}] \label{ex:scrr}
The Randomized Response mechanism described in \cref{example:rr2} has sample complexity: 
$$ N \geq \frac{(n-1)}{\alpha n} \Big[ \frac{n}{(e^{\epsilon} - 1)^2} + \frac{2}{e^{\epsilon}-1}\Big] $$
on the Histogram workload.
\end{example}

\cref{ex:scrr} suggests the sample complexity of the randomized response mechanism grows roughly at a linear rate with the domain size $n$.  

\subsection{Lower Bound}
For a given workload, a theoretical lower bound on the achievable error is useful for checking how close to optimal our strategies are. It also can be used to characterize the inherent difficulty of the workload.  In this section, we derive an easily-computable lower bound on the achievable error under our mechanism in terms of the \emph{singular values} of the workload matrix.


\begin{restatable}[\small{Lower Bound, Factorization Mechanism}]{theorem}{svdb} \label{thm:lowerbound}
Let $\W$ be a workload matrix and let $\Q$ be any $\epsilon$-LDP strategy matrix.   Then we have:
$$ \frac{1}{\exp{(\epsilon)}} (\lambda_1 + \dots + \lambda_n)^2 \leq L(\Q) $$
where $\lambda_1, \dots, \lambda_n$ are the singular values of $\W$ and $L(\Q)$ is the loss function defined in~\cref{thm:objfunQ}.
\end{restatable}
This result is similar to lower bounds known to hold in the central model of differential privacy, based on the analysis of the Matrix Mechanism~\cite{li2015lower}. In both cases the hardness of a workload is characterized by its singular values. 

Other lower bounds for this problem have characterized the hardness of a workload in terms of quantities like the largest $L_2$ column norm of $\W$ \cite{bassily2018linear}, the so-called factorization norm of $\W$ \cite{edmonds2019power}, and the so-called packing number associated with $\W$ \cite{blasiok2019towards}.  While interesting theoretically, the factorization norm and packing number are hard to calculate in practice.  In contrast, our bound can be easily calculated. 

\cref{thm:lowerbound} gives a lower bound on our optimization objective.  Translating that back to worst-case variance gives us \cref{cor:svdbworst}.

\begin{restatable}[Worst-case variance]{corollary}{svdbworst} \label{cor:svdbworst}
The worst-case variance of any factorization mechanism must be at least:
$$ \frac{N}{n \exp{(\epsilon)}} (\lambda_1 + \dots + \lambda_n)^2 - \frac{N}{n} \norm{\W}_F^2 \leq L_{worst}(\V,\Q) $$
\end{restatable}

Combining \cref{cor:svdbworst} with \cref{cor:sample-complexity} and applying it to the Histogram workload gives us a lower bound on the sample complexity.  

\begin{example}[Lower Bound for Histogram Workload] \label{ex:lowerbound}
Every workload factorization mechanism requires at least $ \frac{1}{\alpha} \big( \frac{1}{\exp{(\epsilon)}} - \frac{1}{n}) $ samples to achieve normalized variance $ \alpha$ on the Histogram workload.
\end{example}

Note the very weak dependence on $n$ in \cref{ex:lowerbound}, which suggests that the sample complexity should not change much with $n$.  Further, recall from \cref{ex:scrr} that the sample complexity of randomized response is linear in $n$.  This suggests randomized response is not the best mechanism for the Histogram workload.  This result is not new, as there are several mechanisms that are known to perform better than randomized response \cite{acharya2018,ye2018optimal,wang2017locally,erlingsson2014rappor}.  We show empirically in \cref{sec:experiments} that some of these mechanisms achieve the optimal sample complexity for the Histogram workload up to constant factors (i.e., no dependence on $n$).  Our mechanism also achieves the optimal sample complexity for this workload, but has better constant factors.

For other workloads, the sample complexity may depend on $n$.  Calculating the exact dependence on $n$ for other workloads requires deriving the singular values of the workload as a function of $n$ in closed form, which may be challenging for workloads with complicated structure.

\begin{figure*}[t]
\centering
\includegraphics[width=0.34\textwidth]{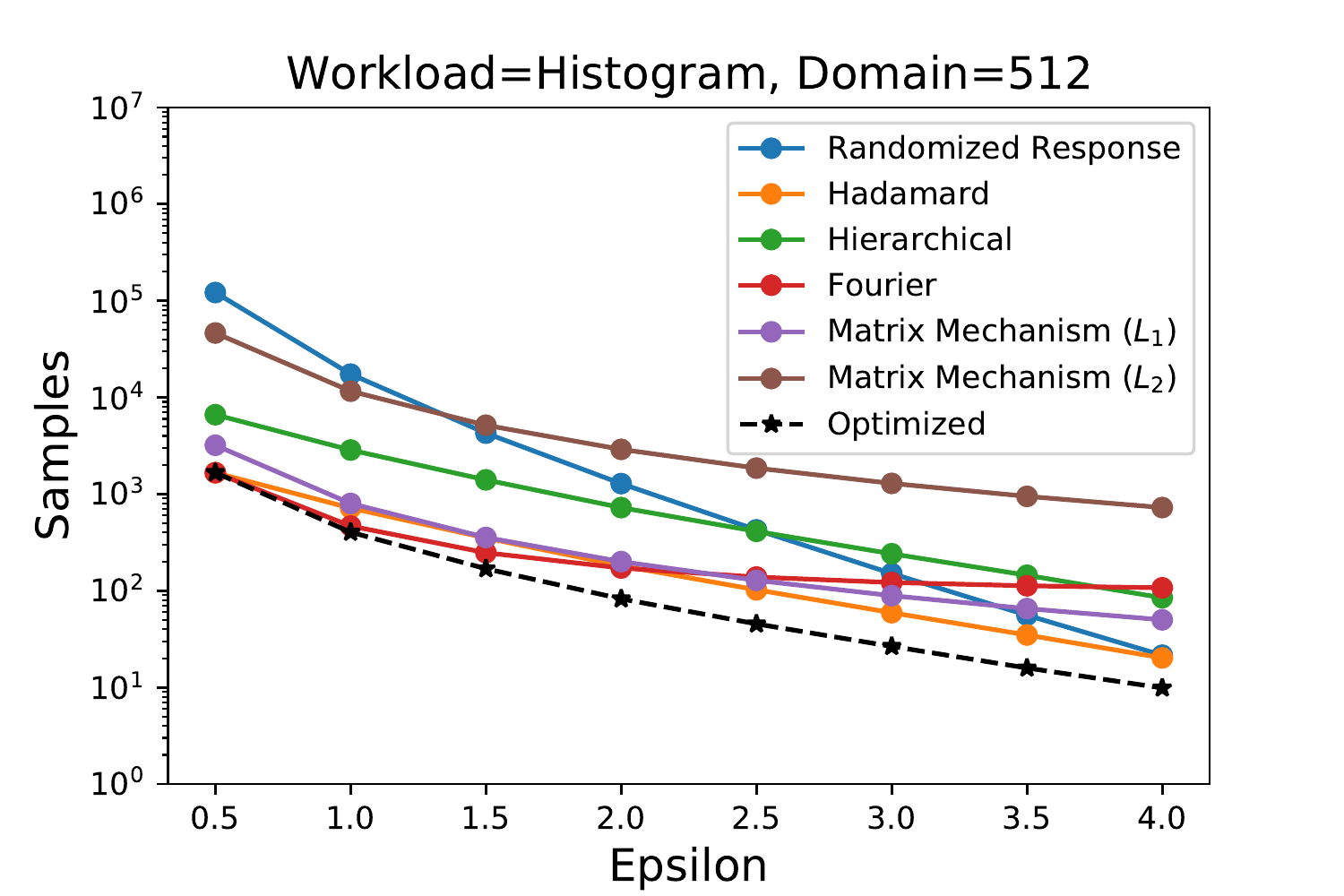} \hspace{-2em}
\includegraphics[width=0.34\textwidth]{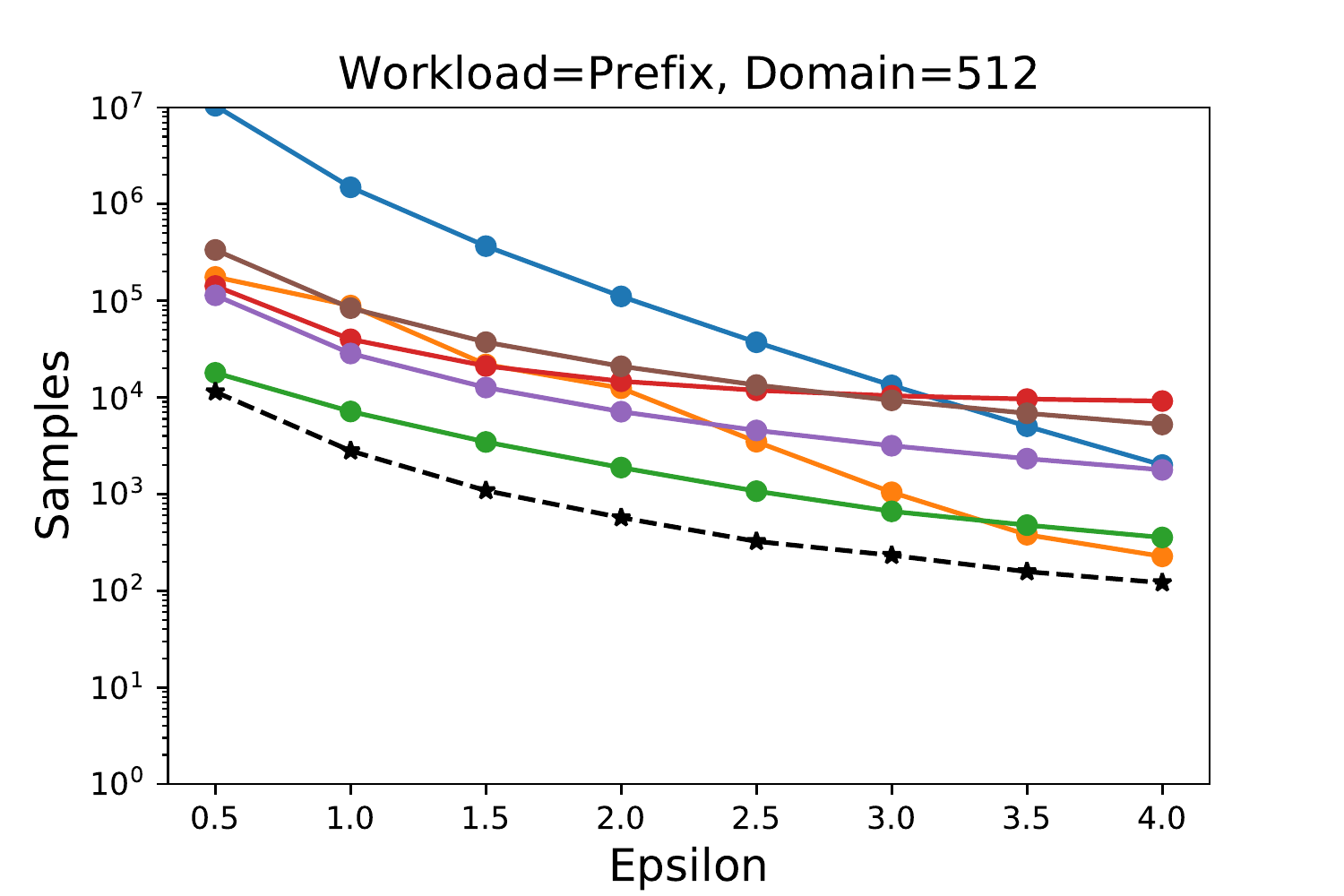}\hspace{-2em}
\includegraphics[width=0.34\textwidth]{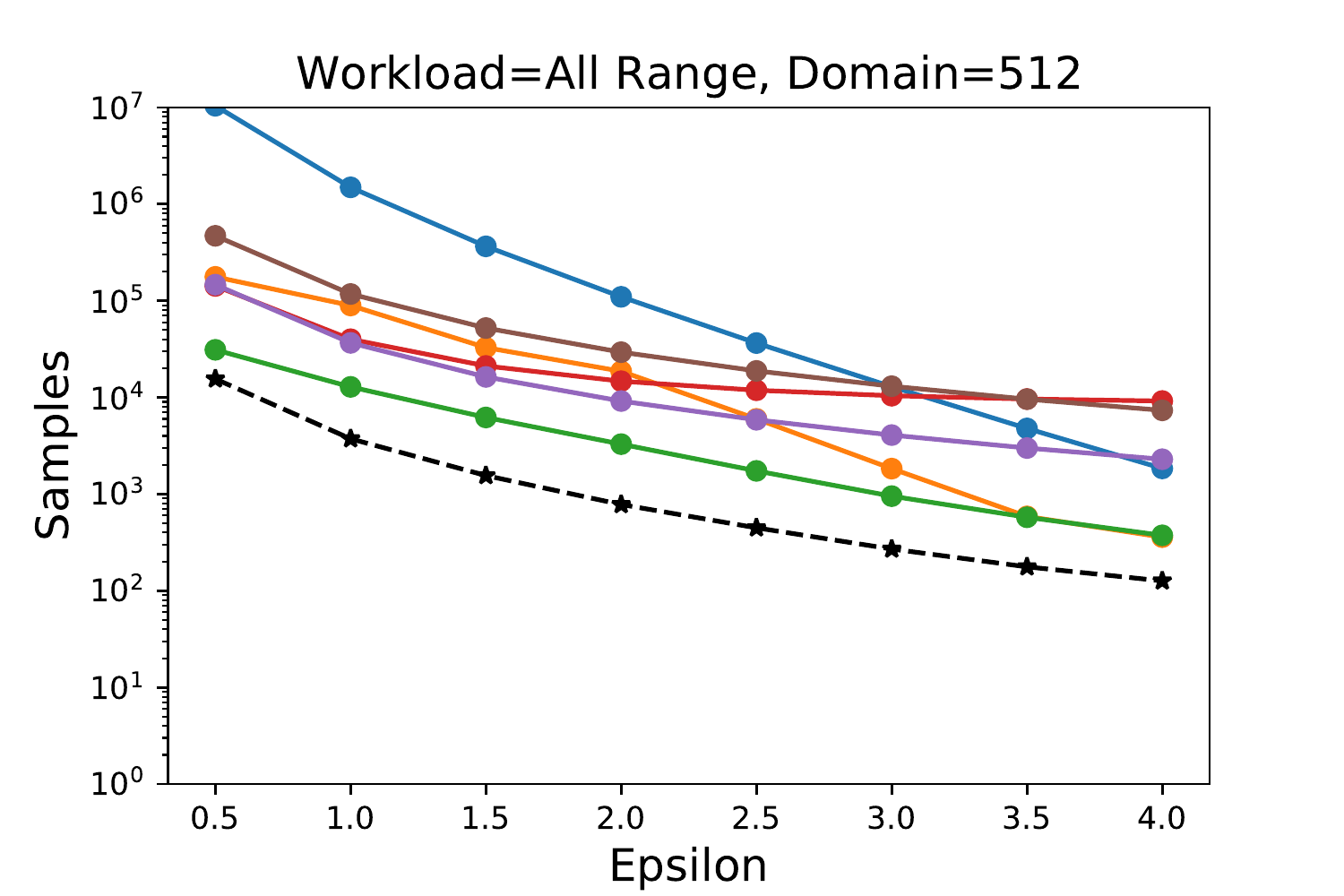}\hspace{-2em}
\includegraphics[width=0.34\textwidth]{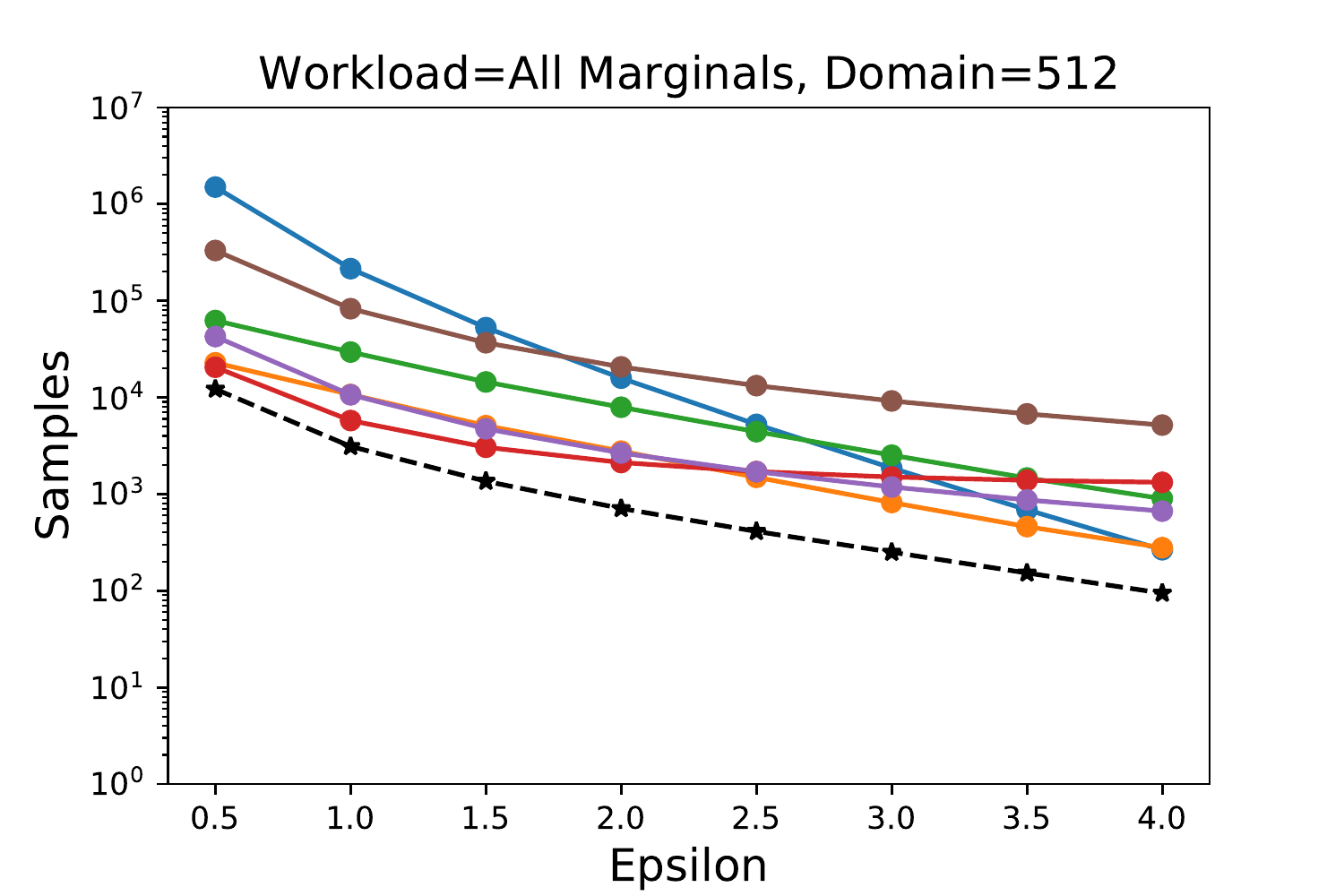}\hspace{-2em}
\includegraphics[width=0.34\textwidth]{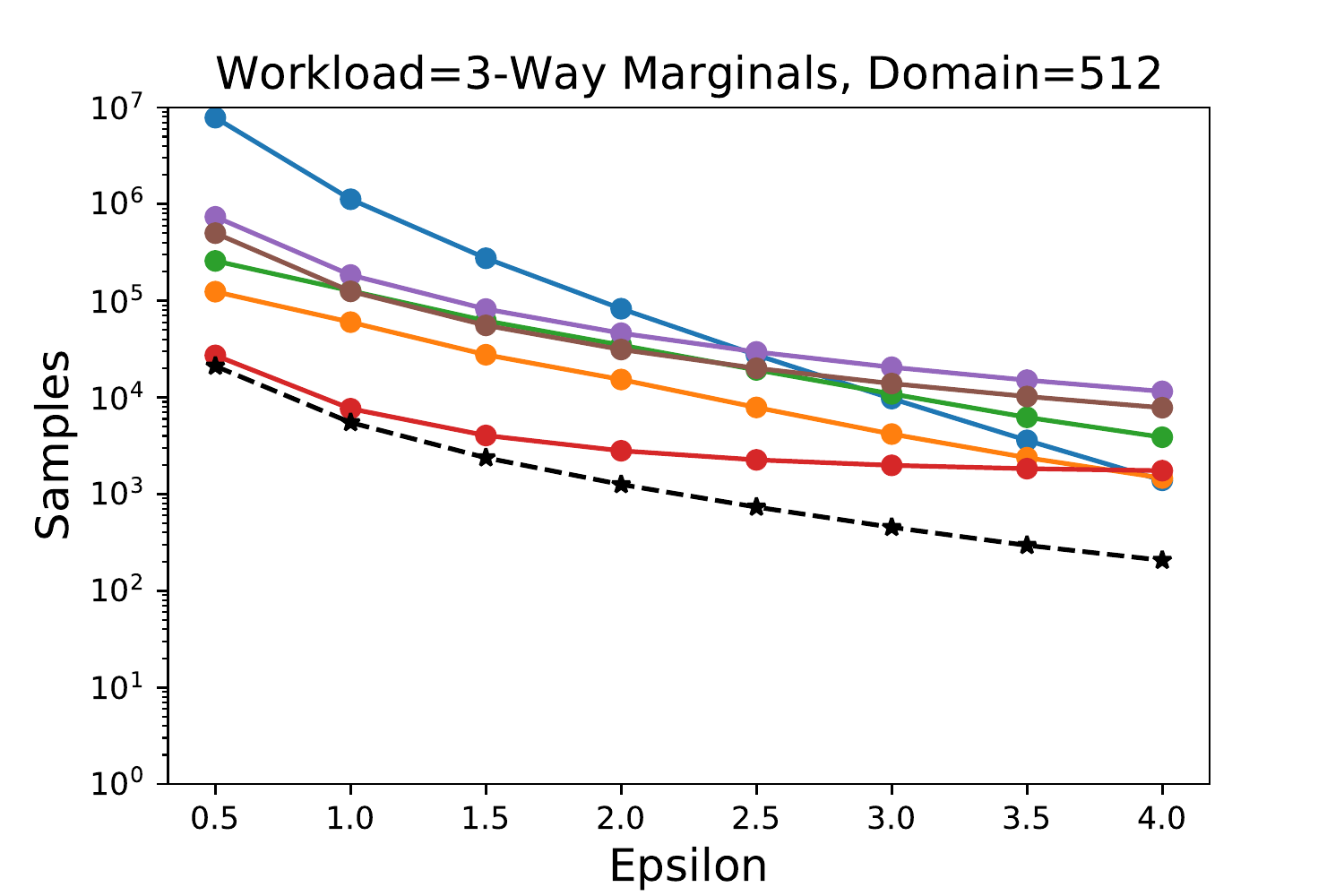}\hspace{-2em}
\includegraphics[width=0.34\textwidth]{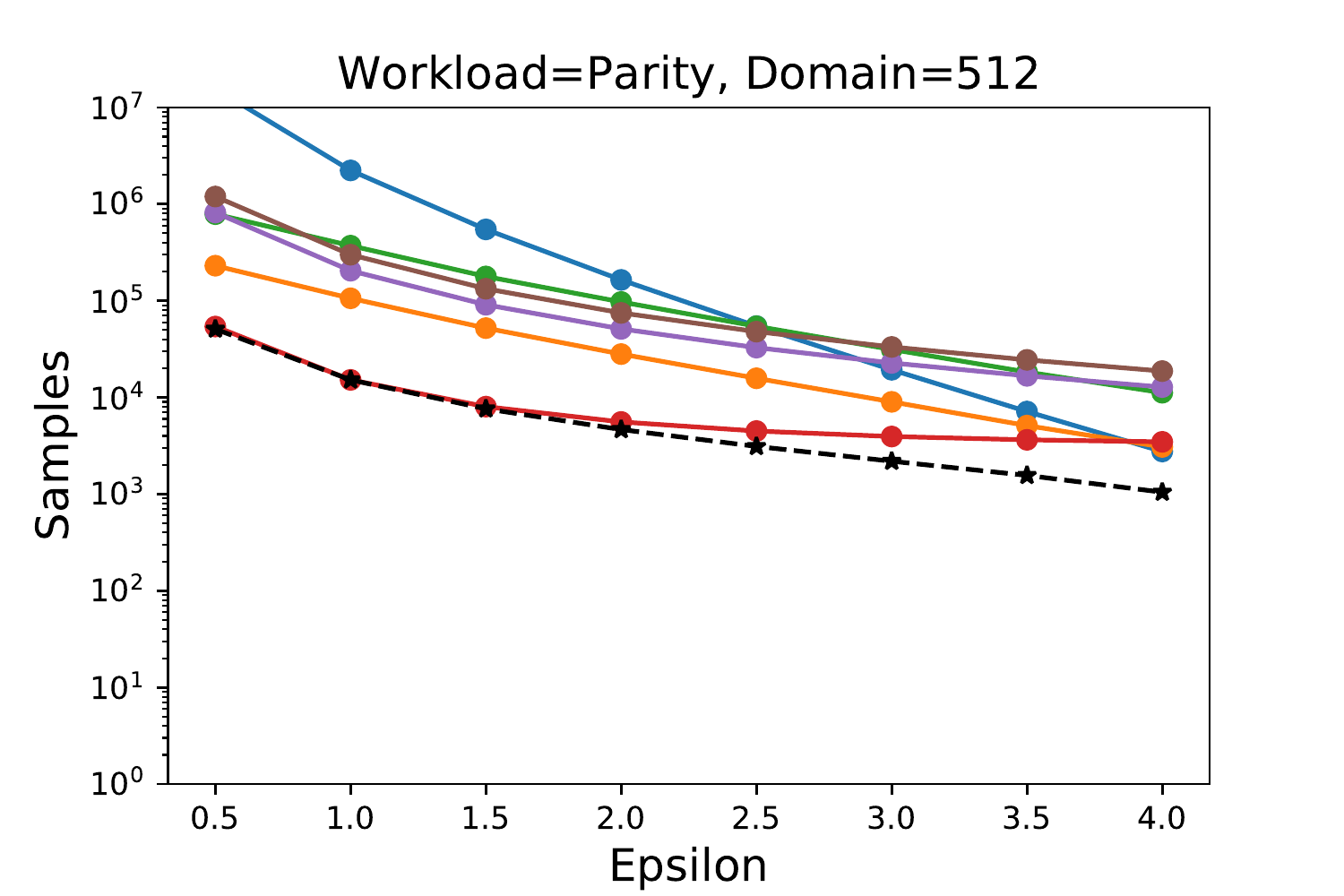}
\caption{ \label{fig:workload-eps} Sample complexity of $7$ algorithms on $6$ workloads for $\epsilon \in [0.5, 4.0]$}
\end{figure*}

\section{Experiments} \label{sec:experiments}

In this section we experimentally evaluate our mechanism.  We extensively study the utility of our mechanism on a variety of workloads, domains, and privacy levels, and compare it against multiple competing mechanisms from the literature.  We demonstrate consistent improvements in utility compared to other mechanisms in all settings (\cref{sec:compare_mech} and \cref{sec:exp:domain} and \cref{sec:exp:dataset}).  We also study the robustness and scalability of our optimization algorithm (\cref{sec:initialization} and \cref{sec:exp:scalability}).  \revision{The source code for our mechanism, and other mechanisms represented as a strategy matrix, is available at \url{https://github.com/ryan112358/workload-factorization-mechanism}.}

\subsection{Experimental setup}
\paragraph*{Workloads}
We consider six different workloads in our empirical analysis, each of which can be defined for a specified domain size.  These workloads are intended to capture common queries an analyst might want to perform on data and have been studied previously in the privacy literature.  These workloads are Histogram, Prefix, All Range, All Marginals, 3-Way Marginals, and Parity.  Histogram is the simplest workload, studied as a running example throughout the paper, and encoded as an identity matrix.  Prefix was introduced in \cref{ex:prefix}, and includes a set of range queries required to compute the empirical CDF over a 1-dimensional domain.  All Range is a workload containing all range queries over a 1-dimensional domain, studied in \cite{cormode2018marginal}.  All Marginals and 3-Way Marginals contain queries to compute the marginals over a multidimensional binary domain, and were studied in \cite{cormode2019answering}.  Parity also contains queries defined over a multidimensional binary domain, and was studied in \cite{gaboardi2014dual}. 

\paragraph*{Mechanisms}

We compare our mechanism against six other state-of-the-art mechanisms, including Randomized Response \cite{warner1965randomized}, Hadamard \cite{acharya2018hadamard}, Hierarchical \cite{cormode2019answering,wang2019answering}, Fourier \cite{cormode2018marginal}, and the Matrix Mechanism \cite{li2010optimizing,edmonds2019power} (both $L_1$ and $L_2$ versions). 
\revision{ 
While the Matrix Mechanism is typically thought of as a technique for central differential privacy, it has been studied theoretically as a mechanism for local differential privacy as well \cite{edmonds2019power}.  This version of the ``distributed'' Matrix Mechanism is what we compare against in experiments.}

The first four mechanisms are all particular instances of the class of factorization mechanisms, just with different factorizations.  They were all designed to answer a fixed workload (e.g., Randomized Response was designed for the Histogram workload), but they can still be run on other workloads with minor modifications.  In particular, for each mechanism we use the same $\Q$ across different workloads, but change $\V$ based on the workload, using \cref{thm:optv}.

We omit from comparison the Gaussian mechanism \cite{bassily2018linear}, as it is strictly dominated by the $L_2$ Matrix Mechanism.  We also omit from comparison RAPPOR \cite{erlingsson2014rappor} and Subset Selection \cite{ye2018optimal}, as they require exponential space to represent the strategy matrix, making it prohibitive to calculate worst-case variance and sample complexity. However, we note that these mechanisms have been previously compared with Hadamard, and shown to offer comparable performance on the Histogram workload \cite{acharya2018hadamard}.


\paragraph*{Evaluation}
Our primary evaluation metric for comparing algorithms is {\em sample complexity}, which we calculate exactly using \cref{cor:sample-complexity} with $\alpha=0.01$.  Recall that the sample complexity is proportional to the worst-case variance, but is appropriately normalized and easier to interpret.  Furthermore, we remark that for most experiments in this section, no input data is required, as the sample complexities we report apply for the worst-case dataset.  In practice, we have found that the variance on real world datasets is quite close to the worst-case variance, as we demonstrate in \cref{sec:exp:dataset}.  We also vary the privacy budget $\epsilon$ and domain size $n$, studying their impact on the sample complexity for each mechanism and workload.

\begin{figure*}[t]
\centering
\includegraphics[width=0.34\textwidth]{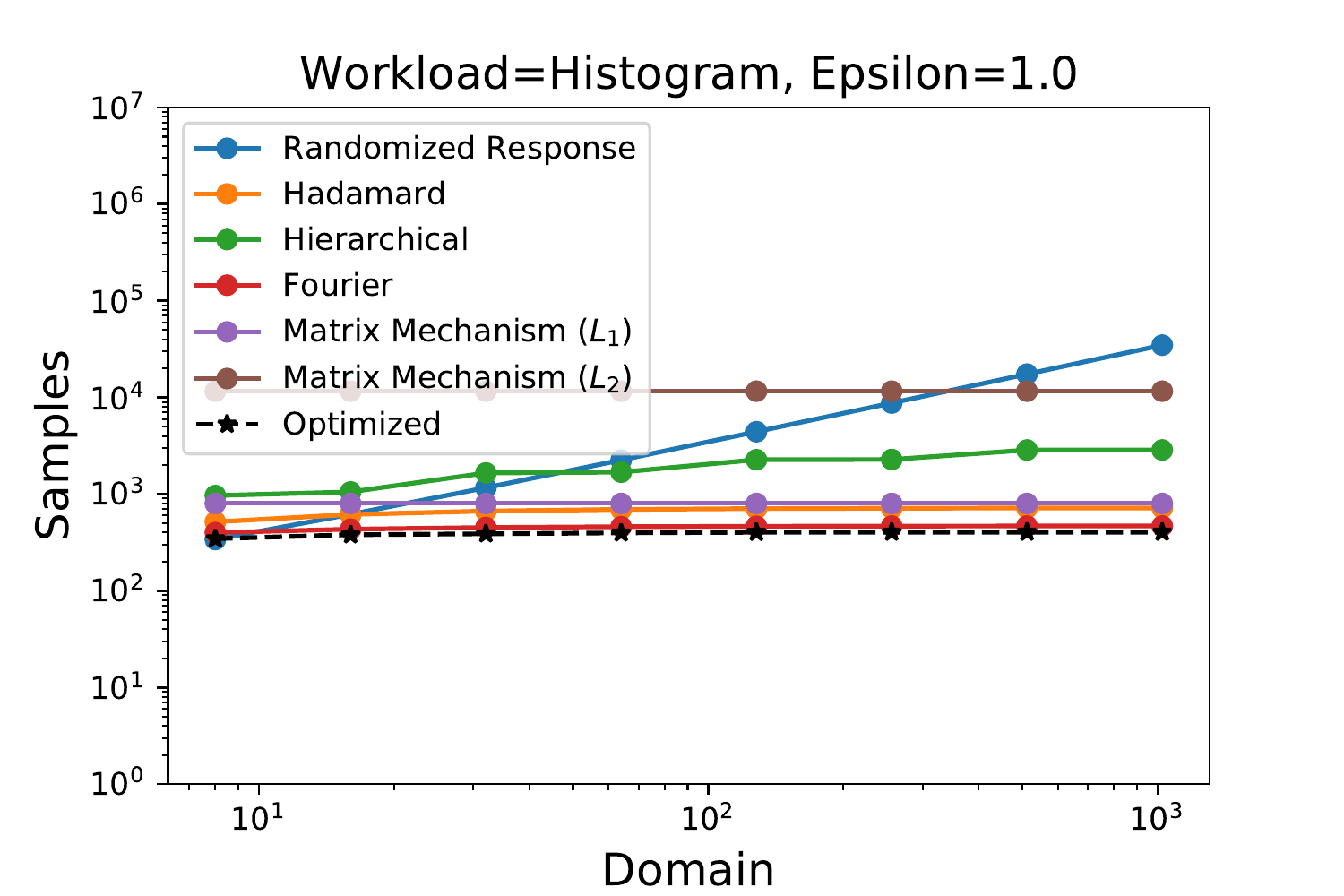} \hspace{-2em}
\includegraphics[width=0.34\textwidth]{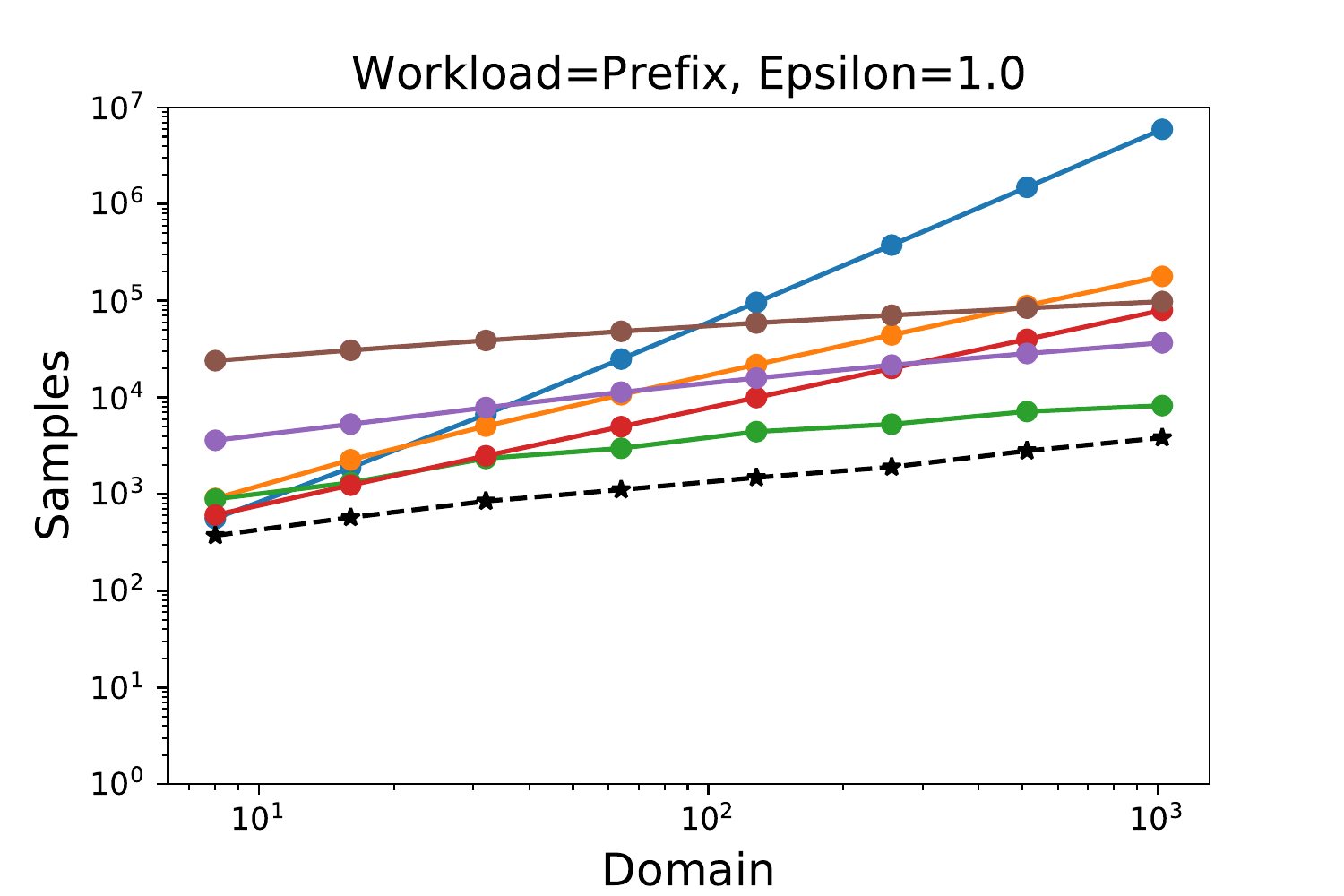}\hspace{-2em}
\includegraphics[width=0.34\textwidth]{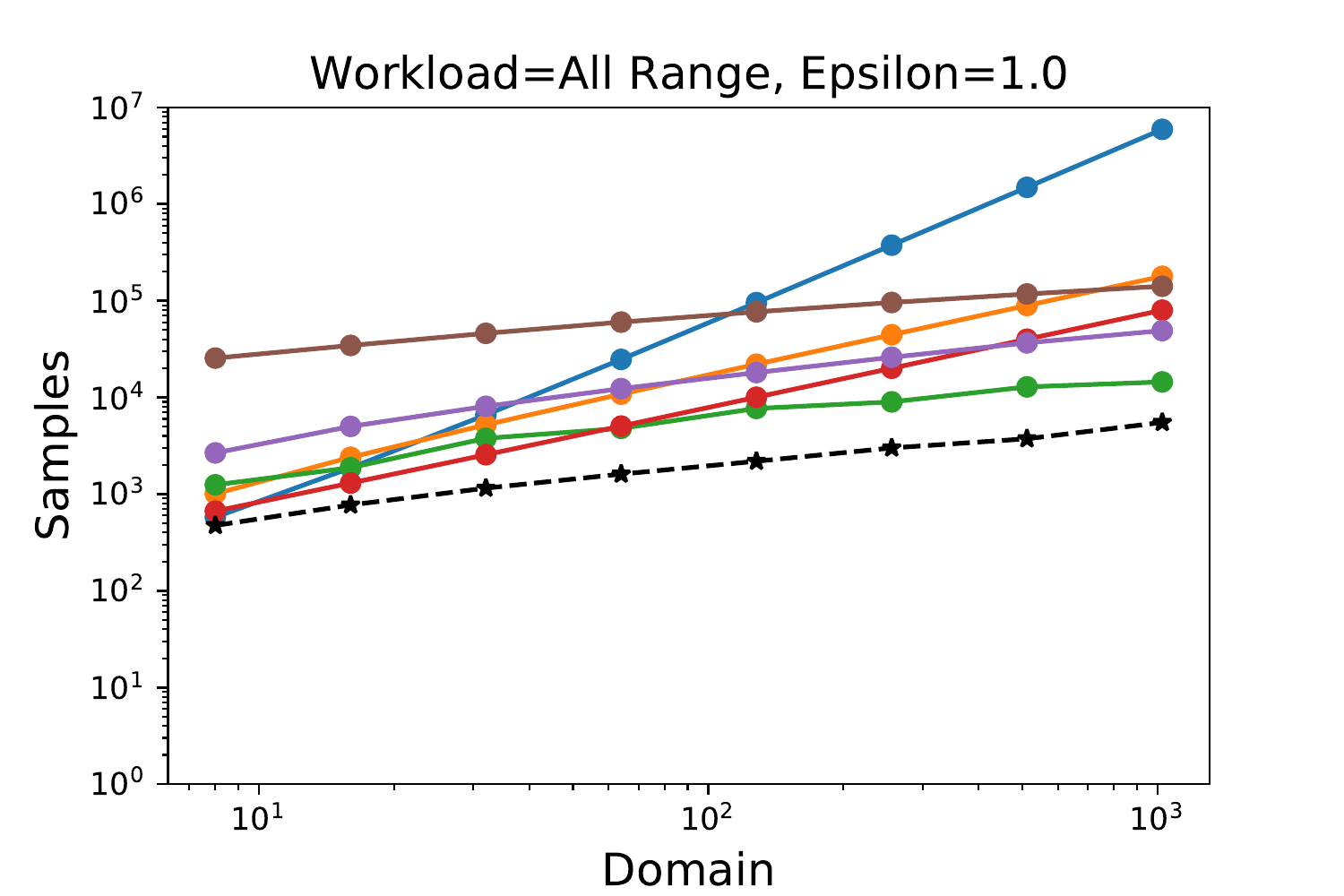}\hspace{-2em}
\includegraphics[width=0.34\textwidth]{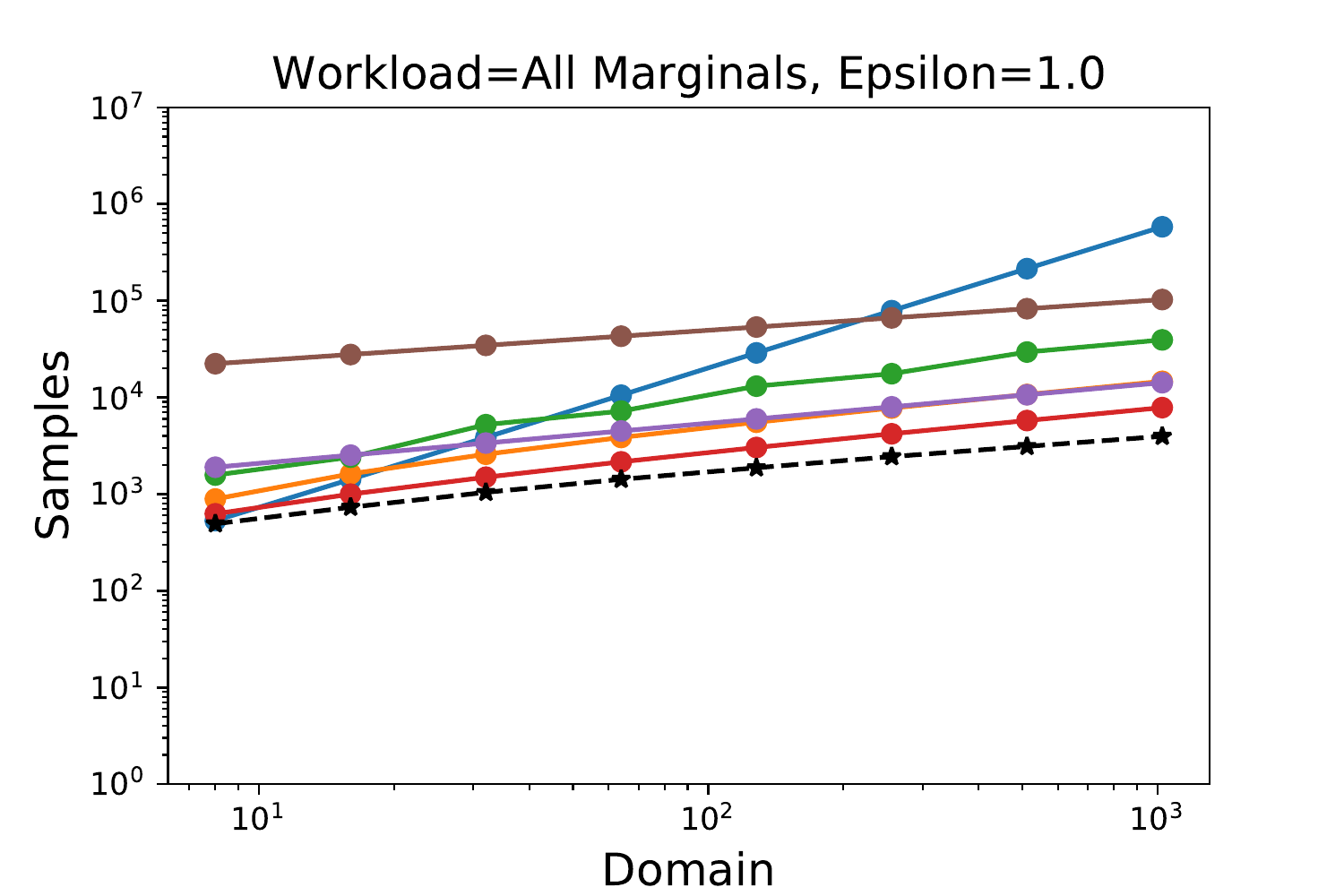}\hspace{-2em}
\includegraphics[width=0.34\textwidth]{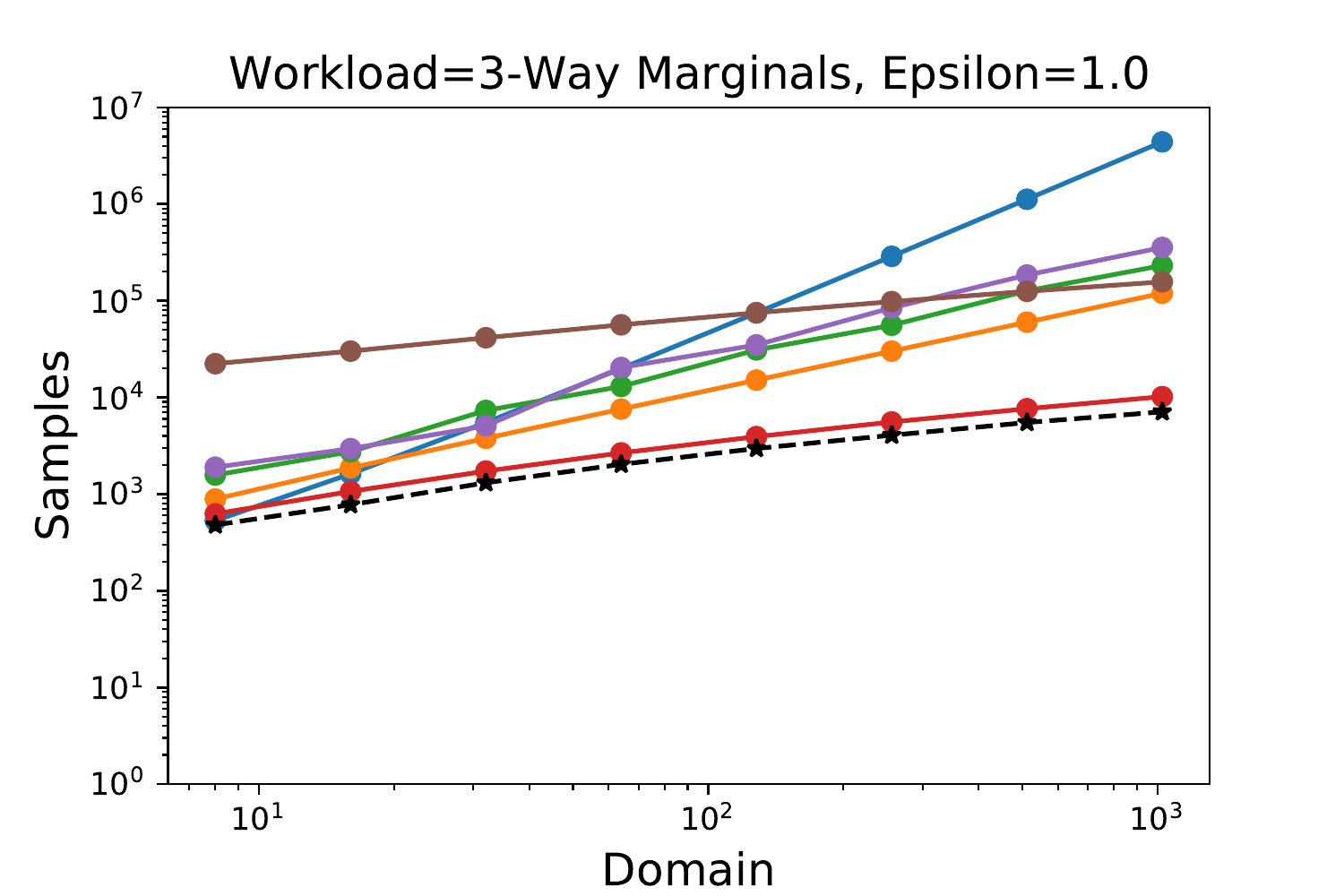}\hspace{-2em}
\includegraphics[width=0.34\textwidth]{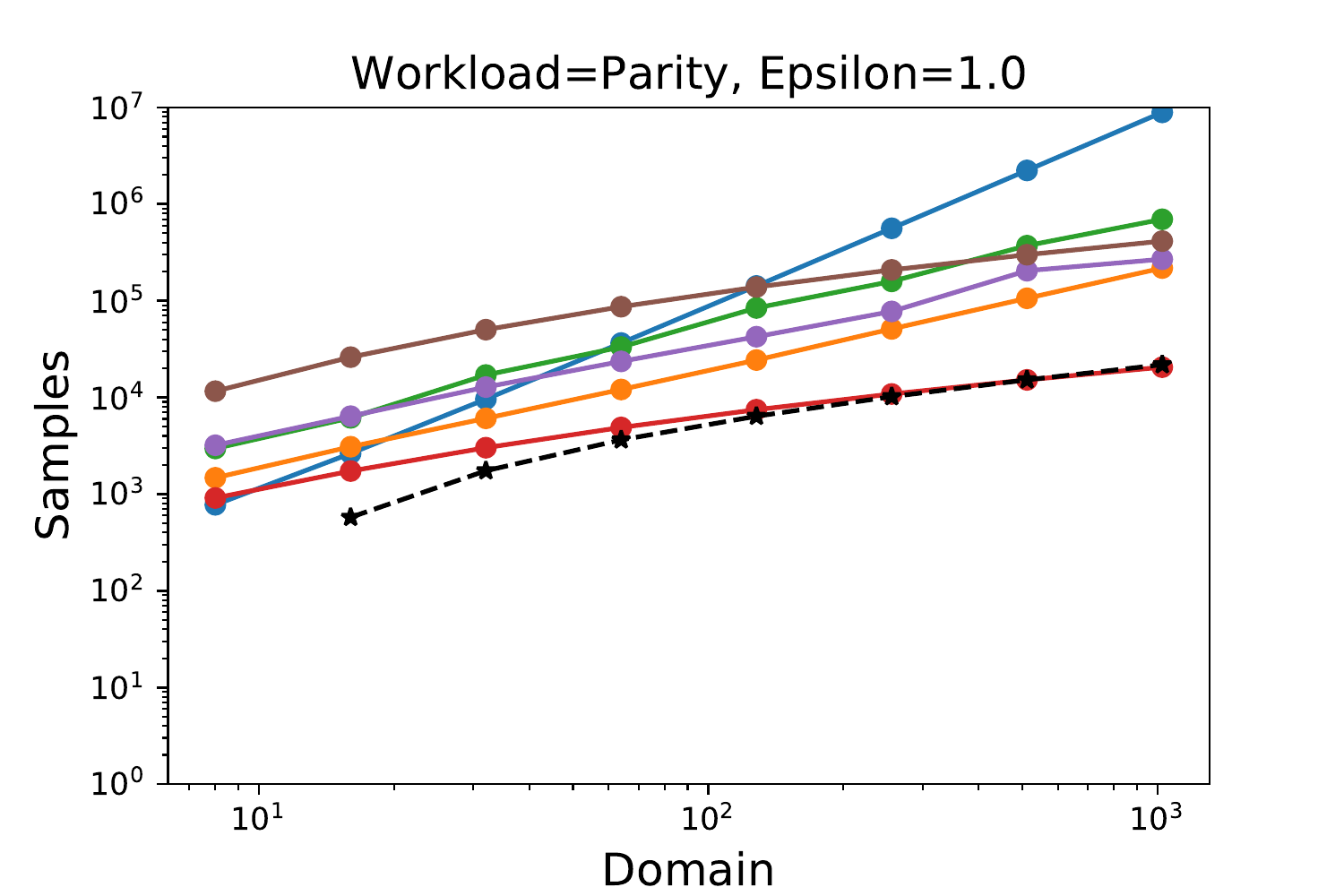}
\caption{ \label{fig:workload-domain} Sample complexity of $7$ algorithms on $6$ workloads for $n \in [8, 1024]$}
\end{figure*}

\subsection{Impact of Epsilon} \label{sec:compare_mech}


Figure \ref{fig:workload-eps} shows the relationship between workload and $\epsilon$ on the sample complexity for each mechanism.  We consider $\epsilon$ ranging from $0.5$ to $4.0$, fixing $n$ to be $512$.  These privacy budgets are common in practical deployments of differential privacy, and local differential privacy in particular \cite{wang2017locally,ding2017collecting,erlingsson2014rappor}.  We state our main findings below:


\begin{itemize}[leftmargin=1em]
\itemsep0em 
\item Our mechanism (labeled Optimized) is consistently the best in all settings: it requires fewer samples than every other mechanism on every workload and $\epsilon$ we tested. 

\item The magnitude of improvement over the best competitor varies between $1.0$ (Histogram, $\epsilon=0.5$) and $14.6$ (All Range, $\epsilon = 4.0$), but the improvement is typically around $2.5$ in the medium privacy regime.  In the very high-privacy regime with $\epsilon \ll 0.5$, our mechanism is typically quite close to the best competitor, and in the very low-privacy regime with $\epsilon \gg 4.0$, our mechanism matches randomized response, which is optimal in that regime.  The reduction of required samples translates to a context that really matters: data collectors can now run their analyses on smaller samples to achieve their desired accuracy.  
\item The best competitor changes with the workload and $\epsilon$.  For example, the best competitor on the Prefix workload was Hierarchical, while the best competitor on the 3-Way Marginals workload was Fourier.  In both cases, these mechanisms were specifically designed to offer low error on their respective workloads, but they don't work as well on other workloads.  
Additionally, Randomized Response is often the best mechanism at high $\epsilon$, so even for a fixed workload, the best competitor is not always the same.
On the other hand, our mechanism adapts effectively to the workload and $\epsilon$, and works well in all settings.  
As a result, only one algorithm needs to be implemented, rather than an entire library of algorithms, and, accordingly it is not necessary to select among alternative algorithms. 
\item Some workloads are inherently harder to answer than others: the number of samples required by our mechanism differs by up to two orders of magnitude between workloads.  The easiest workload appears to be Histogram, while the hardest is Parity.  This is consistent with the lower bound we gave in \cref{thm:lowerbound}, which characterizes the hardness of the workload in terms of its singular values -- the bound is much lower for Histogram than for Parity. 
\end{itemize} 


\begin{figure*}%
    \centering
    \subfloat[\label{fig:dataset}]{{\includegraphics[width=0.33\textwidth]{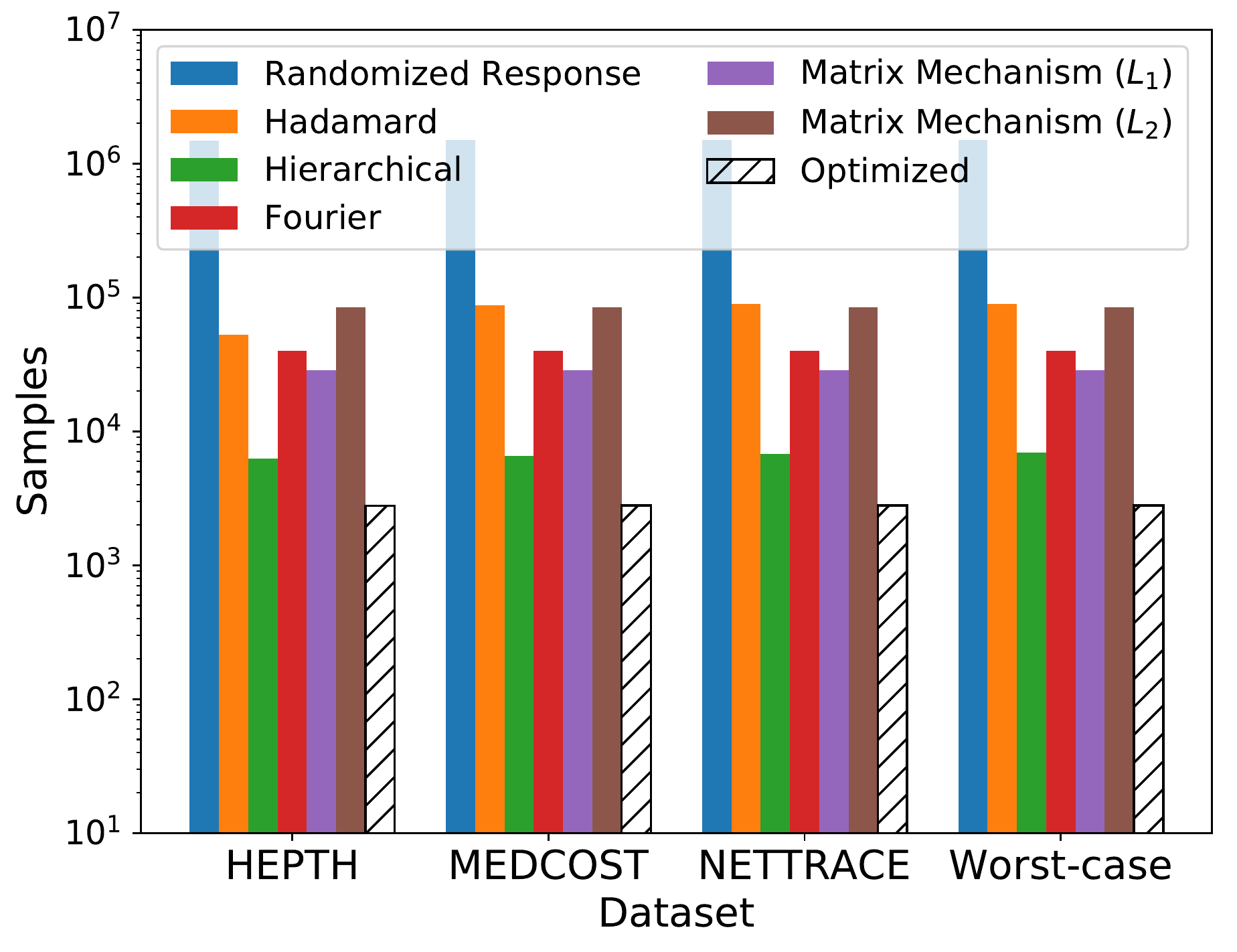}}} 
    \subfloat[\label{fig:initialization}]{{\includegraphics[width=0.33\textwidth]{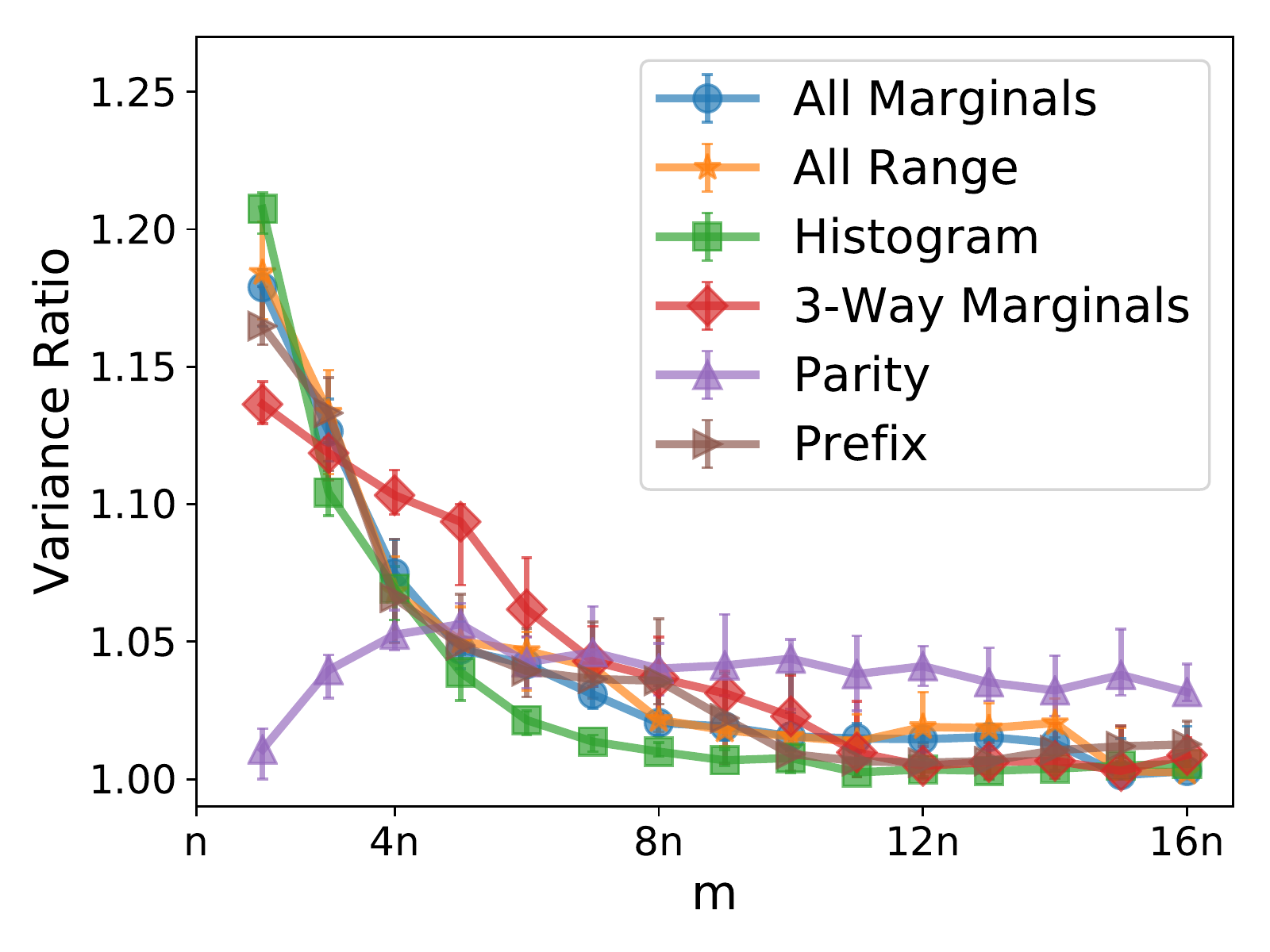} }}%
    \subfloat[ \label{fig:scalability}]{{\includegraphics[width=0.33\textwidth]{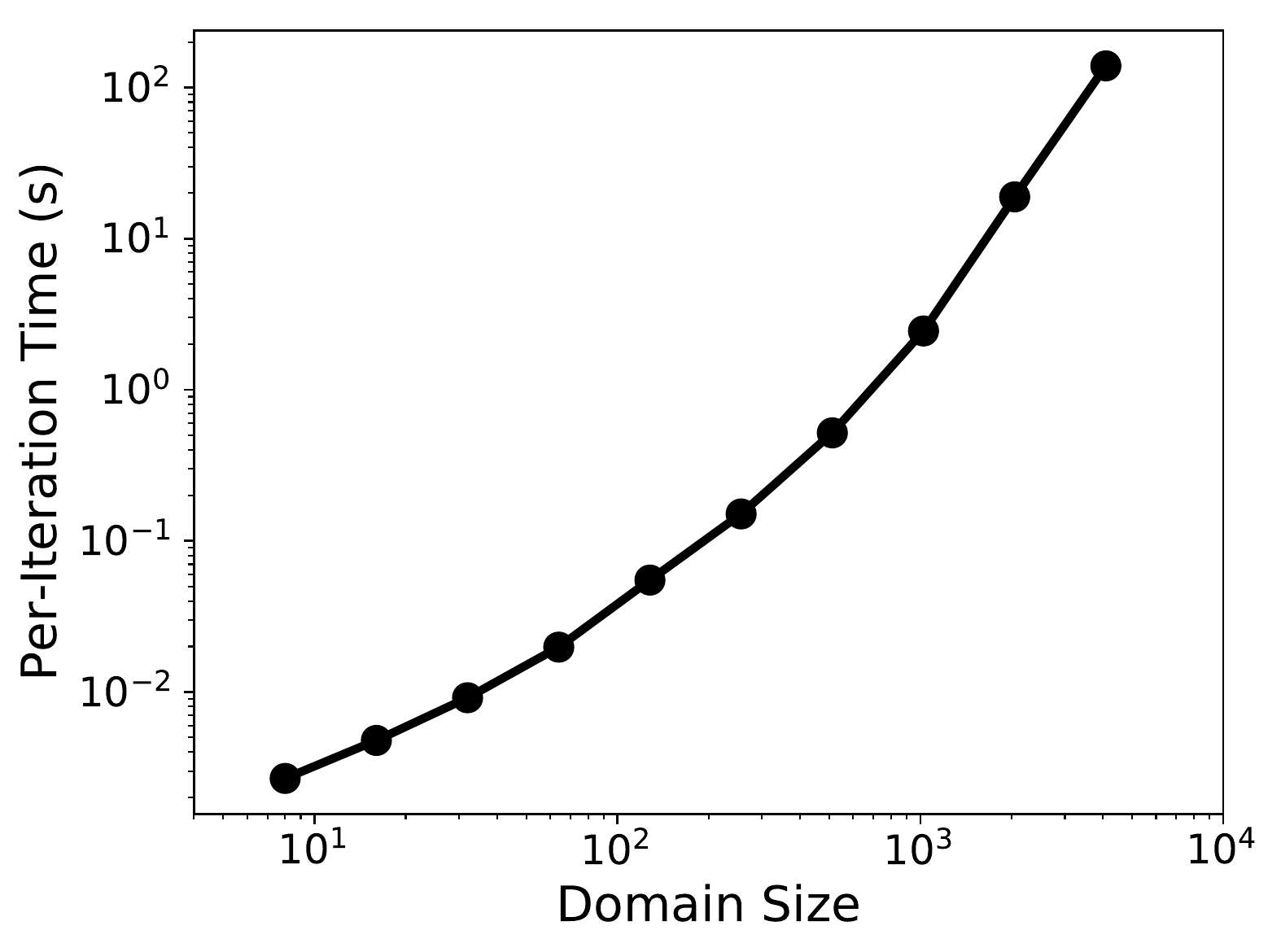} }}%
    \caption{ 
(a) Sample complexity on benchmark datasets for Prefix workload.  
(b) Worst-case variance (ratio to best found) of optimized strategy for various $m$. 
(c) Per-iteration time complexity of optimization for increasing domain sizes. 
}%
\vspace{-1em}
\end{figure*}

\subsection{Impact of Domain Size} \label{sec:exp:domain}

\cref{fig:workload-domain} shows the relationship between workload and $n$ on the sample complexity for each mechanism.  We consider $n$ ranging from $8$ to $1024$, fixing $\epsilon$ to be $1.0$.  
We state our main findings below:


\begin{itemize}[leftmargin=1em]
\itemsep0em 
\item For the Histogram workload, there is almost no dependence on the domain size for every mechanism except randomized response.  This is consistent with our finding in \cref{ex:lowerbound} regarding the lower bound on sample complexity.  This observation is unique to the Histogram workload, however.
\item The mechanisms that were designed for a given workload, and those that adapt to the workload, have a better dependence on the domain size (smaller slope) than the mechanisms that do not.  This includes the $L_2$ Matrix Mechanism, which is worse than every other mechanism in most settings, but slowly overtakes the other mechanisms for large domain sizes.
\item The sample complexity of our mechanism and other mechanisms tailored to the workload is generally $O(\sqrt{n})$, as the slope of the lines are $\approx 0.5$ in log space.  
\footnote{the slope of a line in log space corresponds to the power of a polynomial in linear space; i.e., $\log{y} = m \log{x} \rightarrow y = x^m $.}
On the other hand, the sample complexity of the mechanisms not tailored to the workload is more like $O(n)$ (as the slope of the line is $\approx 1.0$).  These findings are quite interesting: they suggest the improvements offered by workload adaptivity are more than just a constant factor, they grow with the domain size.  

\end{itemize}

\revision{
\subsection{Impact of Dataset} \label{sec:exp:dataset}

Whereas results in previous sections focused on \emph{worst-case} sample complexity, we now turn our attention to sample complexity on \emph{real-world} benchmark datasets obtained from the DPBench study \cite{dpbench}.  To calculate the sample complexity on real data, we simply replace $L_{worst}$ in \cref{cor:sample-complexity} with the exact (data-dependent) expression for total variance stated in \cref{thm:variance}.  

In \cref{fig:dataset}, we plot the sample complexity of each mechanism on three datasets for each mechanism on the Prefix workload, fixing $n=512$ and $\epsilon=1.0$.  We also plot the worst-case sample complexity for reference.  As expected, our mechanism still outperforms all others on each dataset. In fact, all mechanisms performed pretty consistently, offering similar sample complexities for each dataset.  The largest deviation between datasets occurs for the Hadamard mechanism, where it needs $1.69\times$ more samples for the NETTRACE dataset than for the HEPTH dataset.  The Optimized mechanism is even more consistent, as the largest deviation is only $1.006\times$.  Additionally, the real-world sample complexity is very well-approximated by the worst-case sample complexity for the Optimized mechanism, as the maximum deviation is only $1.009\times$.  This suggests the conclusions drawn based on worst-case sample complexity in \cref{fig:workload-eps} and \cref{fig:workload-domain} hold for real-world data as well.  Although not shown, we repeated this experiment for other workloads and settings and made similar observations.
}

\revision{
\subsection{Initialization} \label{sec:initialization}

Recall that our optimization algorithm is initialized with a random strategy matrix, and that different initial strategies can lead to different optimized strategies.  In this section, we aim to understand how sensitive our optimization algorithm is to the different initializations, and whether it depends on $m$, the number of rows in the strategy matrix.  

We fix $n=64$ and $\epsilon=1.0$, and vary $m$ from $2n$ to $16n$.  For each $m$, we compute $10$ optimized strategies with different random initializations and record the worst-case variance for each strategy.  In order to plot all workloads on the same figure, we normalize the worst-case variance to the best found across all trials.  In \cref{fig:initialization}, we plot the median variance ratio for each $m$ as well as an error bar to indicate the min and max ratio obtained across the $10$ trials.  We observe that the optimization is quite robust to the initialization, and produces pretty consistent results between runs, as evident by the small error bars.  Furthermore, the optimization is not very sensitive to the choice of $m$, as all optimized strategies are within a factor of $1.21$ to the best found.  Strategies tend to get closer to optimal for larger $m$, and eventually level off, with the exception of the Parity workload.  We suspect this difference is due to the fact that Parity is a low-rank workload, and doesn't require a large strategy.  Using $m = 4n$ as we did in other experiments tends to produce strategies within a factor of $1.05$ to $1.1$ of the best found.   With enough computational time and resources, we recommend using a hyper-parameter search to find the best $m$, as this extra $10\%$ improvement is meaningful in  practice.  
}

\subsection{Scalability of Optimization} \label{sec:exp:scalability}

We measure the scalability of optimization by looking at the per-iteration time complexity.  In each iteration, we must evaluate the objective function (and its gradient) stated in \cref{thm:objfunQ}, then project onto the constraint set using \cref{alg:projection}.  We assume $\W^T \W$ has been precomputed, and note that the per-iteration time complexity only depends on $\W^T \W$ through its size, and not its contents.  We therefore use the $n\times n$ identity matrix for $\W$.  Additionally, we let $\Q$ be a random $ 4n \times n $ strategy matrix.  In \cref{fig:scalability}, we report the per-iteration time required for increasing domain sizes, averaged over 15 iterations.  As we can see, optimization scales up to domains as large as $n=4096$, where it takes about $139$ seconds per iteration.  While expensive, it is not unreasonable to run for a few hundred iterations in this case, and is an impressive scalability result given that there are over $67$ million optimization variables when $n$ is that large.  Additionally, we note that strategy optimization is a one-time cost, and it can be done offline before deploying the mechanism to the users.  Furthermore, as we showed in \cref{sec:exp:domain}, the number of samples required typically increases with the domain size, so there is good reason to run mechanisms on small domains whenever possible, compressing it if necessary.  In general, the plot shows that the time grows roughly at a $O(n^3)$ rate, as it took about $19$ seconds for $n=2048$ and $2.5$ seconds for $n=1024$, confirming the theoretical time complexity analysis.


\subsection{Non-negativity and consistency} \label{sec:exp:extension}

We now experimentally evaluate the extension we proposed in \cref{remark:extension}, which we call workload non-negative least squares (WNNLS).  The full details are described in \cref{sec:wnnls}.  For this experiment, we fix $\epsilon=1.0$, $N=10^3$, and $n=512$, and use a random sample of data from the ``HEPTH'' dataset obtained from the DPBench study \cite{dpbench}, but note that results on other datasets were similar.  With this extension, we no longer have a closed form expression for variance, so we run $100$ simulations to estimate it instead.

\begin{figure}[h]
\centering
\includegraphics[width=0.33\textwidth]{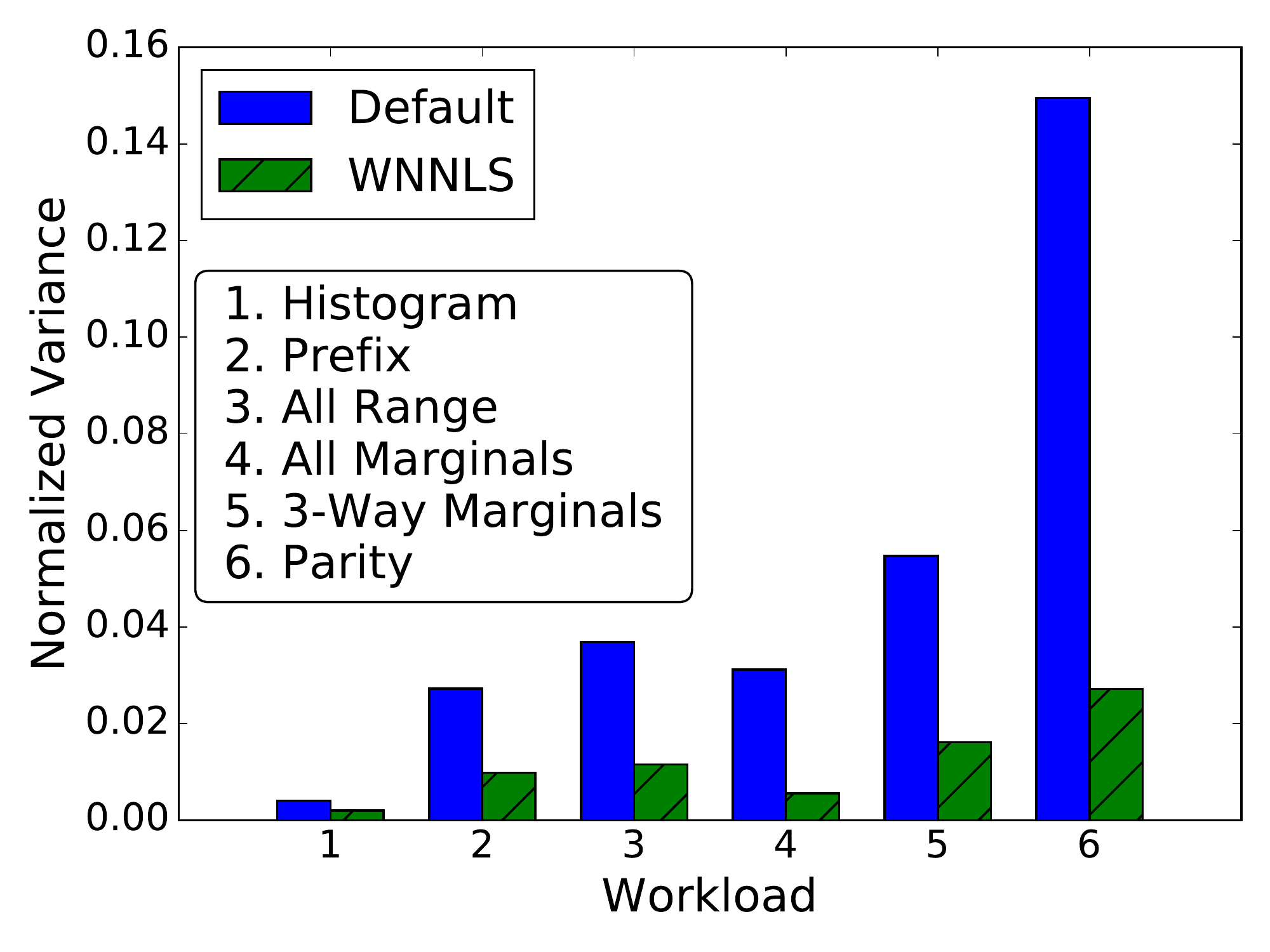}
\vspace{-1em} \caption{ \label{fig:wnnls} Variance of the optimized mechanism with and without the WNNLS extension. }
\end{figure}

Figure \ref{fig:wnnls} shows the (normalized) variance of the mechanism on this dataset with and without the extension.  As we can see, the extension reduces the variance in all cases, and the improvement ranges from $1.96$ to $5.6$, which is a significant amount.  In general, the magnitude of improvement depends on factors like $\epsilon$ and $N$, which are not varied here.  When $N$ and $\epsilon$ are sufficiently large, the default workload query estimates will already be non-negative, in which case WNNLS would offer no improvement.  WNNLS offers significant improvement for many practical $\epsilon$ and $N$, however. 
Additionally, we note that this extension can be plugged into any of the other competing mechanisms as well, and offers similar utility improvements.

\newpage
\section{Related Work} \label{sec:related}

The mechanism we propose in this work is related to a large body of research in both the central and local model of differential privacy.

Answering linear queries under central differential privacy is a widely studied topic \cite{barak2007privacy,li2015matrix,mckenna2018optimizing,bhaskara2012unconditional,li2012adaptive}.  Many state-of-the-art mechanisms for this task, like the Matrix Mechanism \cite{li2010optimizing}, achieve privacy by adding Laplace or Gaussian noise to a carefully selected set of ``strategy queries''.  This query strategy is tailored to the workload, and can even be optimized for it, as the Matrix Mechanism does.  The optimization problem posed by the Matrix Mechanism has been studied theoretically \cite{li2010optimizing,li2015lower}, and several algorithms have been proposed to solve it or approximately solve it \cite{yuan2012low,li2012adaptive,yuan2016convex,mckenna2018optimizing}.  
\revision{
While similar in spirit to our mechanism, the optimization problem underlying the Matrix Mechanism is substantially different from ours, as it requires search over a different space of mechanisms tailored to central differential privacy.  For both optimization problems, the optimization variable is a so-called ``strategy matrix'', but these represent fundamentally different things in each mechanism, and hence the constraints on the strategy matrix differ.  For the Matrix Mechanism, the strategy matrix encodes a set of linear queries which will be answered with a noise-addition mechanism.  In contrast, the strategy matrix for our mechanism encodes a conditional probability distribution.}

\revision{Answering linear queries under local differential privacy has received less attention, but one notable idea is to directly apply mechanisms from central differential privacy to the local model.  This translation can be achieved by simply executing the mechanism independently for each single-user database, then aggregating the results. 
This approach has been studied theoretically with the Gaussian mechanism \cite{bassily2018linear} and the Matrix Mechanism \cite{edmonds2019power}.  While these mechanisms trivially provide privacy, they tend to have poor utility in practice, as they are not tailored to the local model.}
Another notable approach for this task casts it as a mean estimation problem, and uses LDP mechanisms designed for that \cite{blasiok2019towards}.  These works provide a thorough theoretical treatment of this problem, showing bounds on achieved error, but no practical implementation or evaluation.  

More work has been done to answer specific, fixed workloads of general interest, such as histograms \cite{acharya2018,ye2018optimal,wang2017locally, warner1965randomized,erlingsson2014rappor,kairouz2016discrete,bassily2015local,wang2016mutual}, range queries \cite{cormode2019answering,wang2019answering}, and marginals \cite{cormode2018marginal,wang2019answering}.  
A very nice summary of the computational complexity, sample complexity and communication complexity for the mechanisms designed for the Histogram workload is given in \cite{acharya2018}.  Interestingly, even for the very simple Histogram workload, there are multiple mechanisms because the optimal mechanism is not clear.  This is in stark contrast to the central model of differential privacy, where, for the Histogram workload it is clear that the optimal strategy is the workload itself.  Almost all of these mechanisms are instances of the general class of mechanisms we consider in \cref{def:factorization}, just with different workload factorizations.  The strategy matrices for these workloads were all carefully designed to offer low error on the workloads they were designed for, by exploiting knowledge about those specific workloads.  However, none of these mechanisms perform optimization to choose the strategy matrix, instead it is fixed in advance.

Kairouz et al. propose an optimization-based approach to mechanism design as well \cite{kairouz2014extremal}.  The mechanism they propose is not designed to estimate histograms or workload queries, but for other statistical tasks, namely hypothesis testing and something they call information preservation.
They also consider the class of mechanisms characterized by a strategy matrix (\cref{prop:strategy}), and propose an optimization problem over this space to maximize utility for a given task.  Moreover, for convex and sublinear utility functions, they show that the optimal mechanism is a so-called \emph{extremal mechanism}, and state a linear program to find this optimal mechanism.  Unfortunately, there are $2^n$ optimization variables in this linear program, making it infeasible to solve in practice.  Furthermore, the restriction on the utility function (sublinear, convex) prevents the technique from applying to our setting.

\section{Conclusion}

We proposed a new LDP mechanism that adapts to a workload of linear queries provided by an analyst.  We formulated this as a constrained optimization problem over an expressive class of unbiased LDP mechanisms and proposed a projected gradient descent algorithm to solve this problem.  We showed experimentally that our mechanism outperforms all other existing LDP mechanisms in a variety of settings, even outperforming mechanisms on the workloads for which they were intended.

\section*{Acknowledgements}

We would like to thank Daniel Sheldon for his insights regarding \cref{thm:variance}.  This work was supported by the National Science Foundation under grants CNS-1409143, CCF-1642658, CCF-1618512, TRIPODS-1934846; and by DARPA and SPAWAR under contract N66001-15-C-4067.

\newpage
\appendix
\begin{table}[h]
\centering
\begin{tabular}{|c|l|} \hline
Symbol & Meaning \\\hline
$n$ & domain size \\
$N$ & number of users \\
$\epsilon$ & privacy budget \\
$\mathcal{M}$ & privacy mechanism \\
$\U$ & set of possible users, $|\U| = n$ \\
$\O$ & set of possible outcomes, $|\O| = m$ \\
$\W$ & $p \times n $  workload matrix \\
$\Q$ & $m \times n $ strategy matrix \\
$\V$ & $p \times m $ reconstruction matrix \\
$\x$ & data vector \\
$\y$ & response vector, $\y=\M_{\Q}(\x)$ \\
$\mathbf{1}$ & vector of ones \\
$\D$ & $ \text{Diag}(\Q \mathbf{1})$ \\
$\Q^{\dagger}$ & pseudo-inverse \\
$\q_u$ & $u^{th}$ column of $\Q$ \\
$\v_i$ & $i^{th}$ row of $\V$ \\
$tr[\cdot]$ & trace of a matrix \\
$L_{worst}(\V,\Q)$ & worst-case variance \\
$L_{avg}(\V,\Q)$ & average-case variance \\
$L(\V,\Q)$ & optimization objective \\
$L(\Q)$ & optimization objective for $\Q$ \\
$\text{Diag}(\cdot)$ & diagonal matrix from vector \\
\hline
\end{tabular}
\caption{\label{table:notation} Notation}
\end{table}

\section{Non-negativity \& consistency} \label{sec:wnnls}

In this section, we describe and evaluate a simple extension to our mechanism that can greatly improve its utility in practice.    While our basic mechanism offers unbiased answers to the workload queries, this can come at the cost of certain anomalies.  For example, the estimated answer to a workload query might be negative, even though the true answer could never be negative.  This motivates us to consider an extension where we try to account for the structural constraints we know about data vector.  Specifically, we propose the following non-negative least squares problem to find a non-negative (feasible) $\hat{\x}$ such that $ \W \hat{\x}$ is as close as possible to our unbiased estimate $\V \y$.  The problem at hand is:
\begin{align*}
\hat{\x} = \argmin_{\x \geq 0} \norm{\W \x - \V \y}_2^2
\end{align*}
Once $\hat{\x}$ is obtained, the workload answers can be estimated by computing $\W \hat{\x}$.  While this estimate is not necessarily unbiased, it often has substantially lower variance than $\V \y$ and can be a worthwhile trade-off, particularly in the high-privacy/low-data regime where non-negativity is a bigger issue.
A similar problem was studied theoretically by Nikolov et al. \cite{nikolov2013geometry} and empirically by Li et al. \cite{li2015matrix} in the central model of differential privacy, where it has been shown to offer significant utility improvements.  There are various python implementations to solve the above problem efficiently \cite{scipy,zhang2018ektelo,mckenna19graphical}.  We simply use the limited memory BFGS algorithm from scipy to solve it \cite{liu1989limited,scipy}.  

\section{Missing Proofs}\label{sec:pf}
\begin{proof}[of \cref{thm:variance}]
We begin by deriving the variance for a single query $\v^T \y$ (where $\y = \M_{\Q}(\x)$).  Note that $\y$ is a sum of multinomial random variables $\mathbf{s}_u$ instantiated with \revision{parameters} $ n=x_u $ and $\mathbf{p} = \q_u $, \revision{where $\mathbf{s}_u$ is the response vector for all users of type $u$.  Using the well-known formula for the covariance of a multinomial random variable, the covariance of a sum, and the variance of a linear combination of correlated random variables, we obtain:}
\begin{align*}
\mathrm{Cov}[\mathbf{s}_u] &= x_u (\text{Diag}{(\q_u)} - \q_u \q_u^T) \\
\mathrm{Cov}[\y] &= \sum_{u} x_u (\text{Diag}{(\q_u)} - \q_u \q_u^T) \\
\mathrm{Var}[\v^T \y] &= \v^T \mathrm{Cov}[\y] \v \\
&= \sum_{u} x_u (\v^T \text{Diag}{(\q_u)} \v - \v^T \q_u \q_u^T \v) \\
&= \sum_{u} x_u (\v^T \text{Diag}{(\q_u)} \v - (\v^T \q_u)^2)
\end{align*}
The total variance is obtained by summing over all the rows of $\V$.  This completes the proof.
\end{proof}


\begin{proof}[of \cref{thm:objective}]
We prove the claim by showing the two objectives are the same up to constant additive and multiplicative factors.
\begin{align*}
L_{avg}(\V,\Q) 
&\propto \sum_{u \in \U} \sum_{i=1}^p \v_i^T \text{Diag}(\q_u) \v_i - (\v_i^T \q_u)^2 \\
&= \sum_{u \in \U} \sum_{i=1}^p \v_i^T \text{Diag}(\q_u) \v_i - \norm{\V \Q}_F^2 \\
&= \sum_{i=1}^p \v_i^T \text{Diag}\Big(\sum\nolimits_u \q_u \Big) \v_i - \norm{\W}_F^2 \\
&= tr[\V \D \V^T] - \norm{\W}_F^2
\end{align*}
Since $\W$ is constant, we can drop that term for the purposes of defining the objective function.  This completes the proof.
\end{proof}

%


%
%
%


\begin{proof}[of \cref{thm:optv}]
Observe that we can construct $\V$ one row at a time because there are no interaction terms between $\v_i$ and $\v_j$ for $i \neq j$ in the objective function.  Furthermore, we can optimize $\v_i$ through the following quadratic program.
\begin{align*}
& \underset{\v_i}{\text{minimize}}
& & \v_i^T \D \v_i \\
& \text{subject to}
& & \Q^T \v_i = \w_i
\end{align*}

The above problem is closely related to a standard norm minimization problem, and can be transformed to one by making the substitution $\mathbf{u}_i = \D^{\frac{1}{2}} \v_i$.  The problem becomes:
\begin{align*}
& \underset{\mathbf{u}_i}{\text{minimize}}
& & \norm{\mathbf{u}_i}_2^2 \\
& \text{subject to}
& & (\Q^T \D^{-\frac{1}{2}}) \mathbf{u}_i = \w_i
\end{align*}
This unique solution to this problem is given by $\mathbf{u}_i = (\Q^T \D^{-\frac{1}{2}})^\dag \w_i$ \cite{boyd2004convex}.  Using the Hermitian reduction identity \cite{ben2003generalized} $\X^\dag = \X^T (\X \X^T)^\dag $, we have:
\vspace{-5pt}
\begin{align*}
\v_i &= \D^{-\frac{1}{2}} \mathbf{u}_i \\
&= \D^{-\frac{1}{2}} (\Q^T \D^{-\frac{1}{2}})^\dag \w_i \\
&= \D^{-\frac{1}{2}} \D^{-\frac{1}{2}} \Q (\Q^T \D^{-\frac{1}{2}} \D^{-\frac{1}{2}} \Q)^\dag \w_i \\
&= \D^{-1} \Q (\Q^T \D^{-1} \Q)^\dag \w_i 
\end{align*}
Applying this for all $ i $, we arrive at the desired solution. $ \V = \W (\Q^T \D^{-1} \Q)^{\dag} \Q^T \D^{-1} $
\end{proof}

\begin{proof}[of \cref{thm:objfunQ}]
We plug in the optimal solution for $\V$ as given in \cref{thm:optv} and simplify using linear algebra identities and the cyclic permutation property of trace.

\begin{align*}
L(\Q) &\triangleq \min_{\V} L(\V, \Q) \\
&= tr[\W (\Q^T \D^{-1} \Q)^{\dag} \Q^T \D^{-1} \D \D^{-1} \Q (\Q^T \D^{-1} \Q)^{\dag} \W^T] \\
&= tr[\W (\Q^T \D^{-1} \Q)^{\dag} (\Q^T \D^{-1} \Q) (\Q^T \D^{-1} \Q)^{\dag} \W^T] \\
&= tr[\W (\Q^T \D^{-1} \Q)^{\dag} \W^T] \\
&= tr[(\Q^T \D^{-1} \Q)^{\dag} (\W^T \W)]
\end{align*}
\end{proof}

\begin{proof}[of \cref{thm:avg_worst}]

It is obvious that the worst-case variance is greater than (or equal to) the average-case variance.  We will now bound the worst-case variance from above.
Using $u^*$ to denote the worst-case user, we have the following upper bound on worst-case variance:
\begin{align*}
&L_{worst}(\V,\Q)   \\
=& N\max_{u \in \U} \sum_{i=1}^p \v_i^T \text{Diag}(\q_u) \v_i - (\v_i^T \q_u)^2\\
\stackrel{(a)}{\leq} &  N \sum_{i=1}^p \v_i^T \text{Diag}(\q_{u^*}) \v_i \\
= &\frac{N}{n}\sum_{u \in \U} \sum_{i=1}^p \v_i^T \text{Diag}(\q_{u^*}) \v_i \\
\stackrel{(b)}{\leq} & \frac{e^{\epsilon}N}{n}\sum_{u \in \U} \sum_{i=1}^p \v_i^T \text{Diag}(\q_{u}) \v_i \\
\stackrel{(c)}{=} & e^{\epsilon} \Big(\frac{N}{n}\sum_{u \in \U}  \sum_{i=1}^p \v_i^T \text{Diag}(\q_u) \v_i - (\v_i^T \q_u)^2 \Big) + \frac{e^{\epsilon}N}{n}\|\W\|_F^2 \\
=& e^{\epsilon}\big(L_{avg}(\V,\Q) +\frac{N}{n}||\W||_F^2\big)
\end{align*}

In step (a), we use the fact that $(\v_i^T \q_u)^2$ is non-negative.  In step (b), we apply the fact that $ \q_{u*} \leq \exp{(\epsilon)} \q_u$ for all $u$.  In step (c), we express the bound in terms of $L_{avg}$, adding $0$ in the form of $ \norm{\W}_F^2 - \sum_{u} \sum_i (\v_i^T \q_u)^2  $.  This completes the proof.
\end{proof}

\begin{proof}[of \cref{thm:lowerbound}]
Consider the following optimization problem which is closely related to Problem \ref{prob:strategyopt}.
\begin{align*}
& \underset{\X \succ 0}{\text{minimize}}
& & tr[\X^{-1} (\W^T \W)] \\
& \text{subject to}
& & \X_{uu} \leq 1
\end{align*}
Li et al. derived the SVD bound, which shows the minimum above is at least $ \frac{1}{n} (\lambda_1 + \dots + \lambda_n)^2 $ \cite{li2015lower}.  See also \cite{yuan2016convex}.  Furthermore, if $\X^*$ is the optimal solution and the constraint is replaced with $\X_{uu} \leq c$ then $c \X^{*}$ remains optimal \cite{li2015matrix}, in which case the bound becomes $\frac{1}{n c} (\lambda_1 + \dots + \lambda_n)^2 $.

We will now argue that any feasible solution to our problem can be directly transformed into a feasible solution of the above related problem.  Suppose $\Q$ is a feasible solution to Problem \ref{prob:strategyopt} and let $\X = \Q^T \D^{-1} \Q $.  Note that the objective functions are identical now.  We will argue that $\X_{uu} \leq \frac{\exp{(\epsilon)}}{n} $.
\begin{align*}
\X_{uu} &= \sum_{o} \Q_{ou}^2 \frac{1}{\sum_{u'} \Q_{ou}} \\
&\leq \sum_{o} \Q_{ou}^2 \frac{\exp{(\epsilon)}}{n \Q_{ou}} \\
&= \frac{\exp{(\epsilon)}}{n} \sum_o \Q_{ou} \\
&= \frac{\exp{(\epsilon)}}{n} 
\end{align*}
Thus, we have shown that any solution to Problem \ref{prob:strategyopt} gives rise to a corresponding solution to the above problem.  Thus, the SVD Bound applies and we arrive at the desired result:
$$ \frac{1}{\exp{(\epsilon)}} (\lambda_1 + \dots + \lambda_n)^2 \leq L(\Q) $$
\end{proof}


\begin{proof}[of \cref{cor:svdbworst}]
\begin{align*}
L_{worst}(\V,\Q) &\geq L_{avg}(\V,\Q) \\
&= \frac{N}{n} \big[L(\V,\Q) - \norm{\W}_F^2 \big] \\
&\geq \frac{N}{n} \big[L(\Q) - \norm{\W}_F^2 \big] \\
&\geq \frac{N}{n} \Big[ \frac{1}{\exp{(\epsilon)}} (\lambda_1 + \dots + \lambda_n)^2 - \norm{\W}_F^2 \Big]
\end{align*}
\end{proof}


\clearpage
\bibliography{ref}

\begin{thebibliography}{10}

\bibitem{acharya2018}
J.~Acharya, Z.~Sun, and H.~Zhang.
\newblock Communication efficient, sample optimal, linear time locally private
  discrete distribution estimation.
\newblock {\em arXiv preprint arXiv:1802.04705}, 2018.

\bibitem{acharya2018hadamard}
J.~Acharya, Z.~Sun, and H.~Zhang.
\newblock Hadamard response: Estimating distributions privately, efficiently,
  and with little communication.
\newblock {\em arXiv preprint arXiv:1802.04705}, 2018.

\bibitem{barak2007privacy}
B.~Barak, K.~Chaudhuri, C.~Dwork, S.~Kale, F.~McSherry, and K.~Talwar.
\newblock Privacy, accuracy, and consistency too: a holistic solution to
  contingency table release.
\newblock In {\em Proceedings of the twenty-sixth ACM SIGMOD-SIGACT-SIGART
  symposium on Principles of database systems}, pages 273--282. ACM, 2007.

\bibitem{bassily2018linear}
R.~Bassily.
\newblock Linear queries estimation with local differential privacy.
\newblock In K.~Chaudhuri and M.~Sugiyama, editors, {\em Proceedings of Machine
  Learning Research}, volume~89 of {\em Proceedings of Machine Learning
  Research}, pages 721--729. PMLR, 16--18 Apr 2019.

\bibitem{bassily2017practical}
R.~Bassily, K.~Nissim, U.~Stemmer, and A.~G. Thakurta.
\newblock Practical locally private heavy hitters.
\newblock In {\em Advances in Neural Information Processing Systems}, pages
  2288--2296, 2017.

\bibitem{bassily2015local}
R.~Bassily and A.~Smith.
\newblock Local, private, efficient protocols for succinct histograms.
\newblock In {\em Proceedings of the forty-seventh annual ACM symposium on
  Theory of computing}, pages 127--135, 2015.

\bibitem{ben2003generalized}
A.~Ben-Israel and T.~N. Greville.
\newblock {\em Generalized inverses: theory and applications}, volume~15.
\newblock Springer Science \& Business Media, 2003.

\bibitem{bhaskara2012unconditional}
A.~Bhaskara, D.~Dadush, R.~Krishnaswamy, and K.~Talwar.
\newblock Unconditional differentially private mechanisms for linear queries.
\newblock In {\em Proceedings of the forty-fourth annual ACM symposium on
  Theory of computing}, pages 1269--1284. ACM, 2012.

\bibitem{blasiok2019towards}
J.~Blasiok, M.~Bun, A.~Nikolov, and T.~Steinke.
\newblock Towards instance-optimal private query release.
\newblock In {\em Proceedings of the Thirtieth Annual ACM-SIAM Symposium on
  Discrete Algorithms}, pages 2480--2497. Society for Industrial and Applied
  Mathematics, 2019.

\bibitem{boyd2004convex}
S.~Boyd and L.~Vandenberghe.
\newblock {\em Convex optimization}.
\newblock Cambridge university press, 2004.

\bibitem{casella2002statistical}
G.~Casella and R.~L. Berger.
\newblock {\em Statistical inference}, volume~2.
\newblock Duxbury Pacific Grove, CA, 2002.

\bibitem{cormode2018marginal}
G.~Cormode, T.~Kulkarni, and D.~Srivastava.
\newblock Marginal release under local differential privacy.
\newblock In {\em Proceedings of the 2018 International Conference on
  Management of Data}, pages 131--146. ACM, 2018.

\bibitem{cormode2019answering}
G.~Cormode, T.~Kulkarni, and D.~Srivastava.
\newblock Answering range queries under local differential privacy.
\newblock {\em Proceedings of the VLDB Endowment}, 12(10):1126--1138, 2019.

\bibitem{ding2017collecting}
B.~Ding, J.~Kulkarni, and S.~Yekhanin.
\newblock Collecting telemetry data privately.
\newblock In {\em Advances in Neural Information Processing Systems}, pages
  3571--3580, 2017.

\bibitem{dwork2006calibrating}
C.~Dwork, F.~McSherry, K.~Nissim, and A.~Smith.
\newblock Calibrating noise to sensitivity in private data analysis.
\newblock In {\em Theory of cryptography conference}, pages 265--284. Springer,
  2006.

\bibitem{dwork2014algorithmic}
C.~Dwork, A.~Roth, et~al.
\newblock The algorithmic foundations of differential privacy.
\newblock {\em Foundations and Trends{\textregistered} in Theoretical Computer
  Science}, 9(3--4):211--407, 2014.

\bibitem{edmonds2019power}
A.~Edmonds, A.~Nikolov, and J.~Ullman.
\newblock The power of factorization mechanisms in local and central
  differential privacy.
\newblock {\em arXiv preprint arXiv:1911.08339}, 2019.

\bibitem{erlingsson2014rappor}
{\'U}.~Erlingsson, V.~Pihur, and A.~Korolova.
\newblock Rappor: Randomized aggregatable privacy-preserving ordinal response.
\newblock In {\em Proceedings of the 2014 ACM SIGSAC conference on computer and
  communications security}, pages 1054--1067. ACM, 2014.

\bibitem{gaboardi2014dual}
M.~Gaboardi, E.~J.~G. Arias, J.~Hsu, A.~Roth, and Z.~S. Wu.
\newblock Dual query: Practical private query release for high dimensional
  data.
\newblock In {\em International Conference on Machine Learning}, pages
  1170--1178, 2014.

\bibitem{griewank1989automatic}
A.~Griewank et~al.
\newblock On automatic differentiation.
\newblock {\em Mathematical Programming: recent developments and applications},
  6(6):83--107, 1989.

\bibitem{hardt2012simple}
M.~Hardt, K.~Ligett, and F.~McSherry.
\newblock A simple and practical algorithm for differentially private data
  release.
\newblock In {\em Advances in Neural Information Processing Systems}, pages
  2339--2347, 2012.

\bibitem{dpbench}
M.~Hay, A.~Machanavajjhala, G.~Miklau, Y.~Chen, and D.~Zhang.
\newblock Principled evaluation of differentially private algorithms using
  dpbench.
\newblock In {\em Proceedings of the 2016 International Conference on
  Management of Data}, pages 139--154. ACM, 2016.

\bibitem{hay2010boosting}
M.~Hay, V.~Rastogi, G.~Miklau, and D.~Suciu.
\newblock Boosting the accuracy of differentially private histograms through
  consistency.
\newblock {\em Proceedings of the VLDB Endowment}, 3(1-2):1021--1032, 2010.

\bibitem{holohan2017extreme}
N.~Holohan, D.~J. Leith, and O.~Mason.
\newblock Extreme points of the local differential privacy polytope.
\newblock {\em Linear Algebra and its Applications}, 534:78--96, 2017.

\bibitem{kairouz2016discrete}
P.~Kairouz, K.~Bonawitz, and D.~Ramage.
\newblock Discrete distribution estimation under local privacy.
\newblock {\em arXiv preprint arXiv:1602.07387}, 2016.

\bibitem{kairouz2014extremal}
P.~Kairouz, S.~Oh, and P.~Viswanath.
\newblock Extremal mechanisms for local differential privacy.
\newblock In {\em Advances in neural information processing systems}, pages
  2879--2887, 2014.

\bibitem{li2010optimizing}
C.~Li, M.~Hay, V.~Rastogi, G.~Miklau, and A.~McGregor.
\newblock Optimizing linear counting queries under differential privacy.
\newblock In {\em Proceedings of the twenty-ninth ACM SIGMOD-SIGACT-SIGART
  symposium on Principles of database systems}, pages 123--134. ACM, 2010.

\bibitem{li2012adaptive}
C.~Li and G.~Miklau.
\newblock An adaptive mechanism for accurate query answering under differential
  privacy.
\newblock {\em Proceedings of the VLDB Endowment}, 5(6):514--525, 2012.

\bibitem{li2015lower}
C.~Li and G.~Miklau.
\newblock Lower bounds on the error of query sets under the
  differentially-private matrix mechanism.
\newblock {\em Theory of Computing Systems}, 57(4):1159--1201, 2015.

\bibitem{li2015matrix}
C.~Li, G.~Miklau, M.~Hay, A.~McGregor, and V.~Rastogi.
\newblock The matrix mechanism: optimizing linear counting queries under
  differential privacy.
\newblock {\em The VLDB journal}, 24(6):757--781, 2015.

\bibitem{liu1989limited}
D.~C. Liu and J.~Nocedal.
\newblock On the limited memory bfgs method for large scale optimization.
\newblock {\em Mathematical programming}, 45(1-3):503--528, 1989.

\bibitem{maclaurin2015autograd}
D.~Maclaurin, D.~Duvenaud, and R.~P. Adams.
\newblock Autograd: Effortless gradients in numpy.
\newblock In {\em ICML 2015 AutoML Workshop}, volume 238, 2015.

\bibitem{mckenna2018optimizing}
R.~McKenna, G.~Miklau, M.~Hay, and A.~Machanavajjhala.
\newblock Optimizing error of high-dimensional statistical queries under
  differential privacy.
\newblock {\em Proceedings of the VLDB Endowment}, 11(10):1206--1219, 2018.

\bibitem{mckenna19graphical}
R.~McKenna, D.~Sheldon, and G.~Miklau.
\newblock Graphical-model based estimation and inference for differential
  privacy.
\newblock In {\em Proceedings of the 36th International Conference on Machine
  Learning (ICML)}, 2019.

\bibitem{nikolov2013geometry}
A.~Nikolov, K.~Talwar, and L.~Zhang.
\newblock The geometry of differential privacy: the sparse and approximate
  cases.
\newblock In {\em Proceedings of the forty-fifth annual ACM symposium on Theory
  of computing}, pages 351--360. ACM, 2013.

\bibitem{nocedal2006numerical}
J.~Nocedal and S.~Wright.
\newblock {\em Numerical optimization}.
\newblock Springer Science \& Business Media, 2006.

\bibitem{strang1993introduction}
G.~Strang, G.~Strang, G.~Strang, and G.~Strang.
\newblock {\em Introduction to linear algebra}, volume~3.
\newblock Wellesley-Cambridge Press Wellesley, MA, 1993.

\bibitem{thakurta2017learning}
A.~G. Thakurta, A.~H. Vyrros, U.~S. Vaishampayan, G.~Kapoor, J.~Freudiger,
  V.~R. Sridhar, and D.~Davidson.
\newblock Learning new words, Mar.~14 2017.
\newblock US Patent 9,594,741.

\bibitem{scipy}
P.~{Virtanen}, R.~{Gommers}, T.~E. {Oliphant}, M.~{Haberland}, T.~{Reddy},
  D.~{Cournapeau}, E.~{Burovski}, P.~{Peterson}, W.~{Weckesser}, J.~{Bright},
  S.~J. {van der Walt}, M.~{Brett}, J.~{Wilson}, K.~{Jarrod Millman},
  N.~{Mayorov}, A.~R.~J. {Nelson}, E.~{Jones}, R.~{Kern}, E.~{Larson},
  C.~{Carey}, {\.I}.~{Polat}, Y.~{Feng}, E.~W. {Moore}, J.~{Vand erPlas},
  D.~{Laxalde}, J.~{Perktold}, R.~{Cimrman}, I.~{Henriksen}, E.~A. {Quintero},
  C.~R. {Harris}, A.~M. {Archibald}, A.~H. {Ribeiro}, F.~{Pedregosa}, P.~{van
  Mulbregt}, and S.~.~. {Contributors}.
\newblock {SciPy 1.0--Fundamental Algorithms for Scientific Computing in
  Python}.
\newblock {\em arXiv e-prints}, page arXiv:1907.10121, Jul 2019.

\bibitem{wang2016mutual}
S.~Wang, L.~Huang, P.~Wang, Y.~Nie, H.~Xu, W.~Yang, X.-Y. Li, and C.~Qiao.
\newblock Mutual information optimally local private discrete distribution
  estimation.
\newblock {\em arXiv preprint arXiv:1607.08025}, 2016.

\bibitem{wang2017locally}
T.~Wang, J.~Blocki, N.~Li, and S.~Jha.
\newblock Locally differentially private protocols for frequency estimation.
\newblock In {\em Proc. of the 26th USENIX Security Symposium}, pages 729--745,
  2017.

\bibitem{wang2019answering}
T.~Wang, B.~Ding, J.~Zhou, C.~Hong, Z.~Huang, N.~Li, and S.~Jha.
\newblock Answering multi-dimensional analytical queries under local
  differential privacy.
\newblock In {\em Proceedings of the 2019 International Conference on
  Management of Data}, pages 159--176, 2019.

\bibitem{wang2013projection}
W.~Wang and M.~A. Carreira-Perpin{\'a}n.
\newblock Projection onto the probability simplex: An efficient algorithm with
  a simple proof, and an application.
\newblock {\em arXiv preprint arXiv:1309.1541}, 2013.

\bibitem{warner1965randomized}
S.~L. Warner.
\newblock Randomized response: A survey technique for eliminating evasive
  answer bias.
\newblock {\em Journal of the American Statistical Association},
  60(309):63--69, 1965.

\bibitem{ye2018optimal}
M.~Ye and A.~Barg.
\newblock Optimal schemes for discrete distribution estimation under locally
  differential privacy.
\newblock {\em IEEE Transactions on Information Theory}, 64(8):5662--5676,
  2018.

\bibitem{yuan2016convex}
G.~Yuan, Y.~Yang, Z.~Zhang, and Z.~Hao.
\newblock Convex optimization for linear query processing under approximate
  differential privacy.
\newblock In {\em Proceedings of the 22nd ACM SIGKDD International Conference
  on Knowledge Discovery and Data Mining}, pages 2005--2014. ACM, 2016.

\bibitem{yuan2012low}
G.~Yuan, Z.~Zhang, M.~Winslett, X.~Xiao, Y.~Yang, and Z.~Hao.
\newblock Low-rank mechanism: optimizing batch queries under differential
  privacy.
\newblock {\em Proceedings of the VLDB Endowment}, 5(11):1352--1363, 2012.

\bibitem{zhang2018ektelo}
D.~Zhang, R.~McKenna, I.~Kotsogiannis, M.~Hay, A.~Machanavajjhala, and
  G.~Miklau.
\newblock Ektelo: A framework for defining differentially-private computations.
\newblock In {\em Proceedings of the 2018 International Conference on
  Management of Data}, pages 115--130. ACM, 2018.

\end{thebibliography}
\bibliographystyle{abbrv}

\end{document}